\documentclass[1p]{elsarticle}

\usepackage{hyperref}

\usepackage{epsfig,graphicx,wrapfig,pgf}
\usepackage[utf8]{inputenc}
\usepackage[english]{babel}
\usepackage{amsthm}
\usepackage{caption}
\usepackage{soul}
\usepackage{subcaption}
\usepackage{graphicx} 
\newcommand\tab[1][1cm]{\hspace*{#1}}
\usepackage{filecontents,pgfplots}
\usepgfplotslibrary{external} 
\tikzsetexternalprefix{figs/}
\graphicspath{ {figures/} }
\newtheorem{remark}{Remark}[section]
\usepackage{amsmath}

\DeclareCaptionFormat{subfig}{\figurename~#1#2#3}
\DeclareCaptionSubType*{figure}
\journal{arXiv.org}

\begin{document}

\begin{frontmatter}

\title{An Efficient Algorithm for the Multicomponent Compressible Navier-Stokes Equations in Low- and High-Mach Number Regimes.}

\author{Roman Frolov (frolov@ualberta.ca)\\
© 2018. This manuscript version is made available under the CC-BY-NC-ND 4.0 license http://creativecommons.org/licenses/by-nc-nd/4.0/}



\address[mymainaddress]{Department of Mathematical and Statistical Sciences, University of Alberta, Edmonton, AB, Canada}

\begin{abstract}
The goal of this study is to develop an efficient numerical algorithm applicable to a wide range of compressible multicomponent flows, including nearly incompressible low-Mach number flows, flows with strong shocks, multicomponent flows with high density ratios and interfacial physics, inviscid and viscous flows, as well as flows featuring combinations of these phenomena and various interactions between them. Although many highly efficient algorithms have been proposed for simulating each type of the flows mentioned above, the construction of a universal solver is known to be challenging. Extreme cases, such as incompressible and highly compressible flows, or inviscid and highly viscous flows, require different numerical treatments in order to maintain the efficiency, stability, and accuracy of the method.

Linearized block implicit (LBI) factored schemes (see e.g. \cite{brileystruct}, \cite{beamwarming}) are known to provide an efficient way of solving the compressible Navier-Stokes equations (compressible NSEs) implicitly, allowing us to avoid stability restrictions at low Mach number and high viscosity. However, the methods' splitting error has been shown to grow and dominate physical fluxes as the Mach number approaches zero (see \cite{choi}). In this paper, a splitting error reduction technique is proposed to solve the issue. A shock-capturing algorithm from \cite{invdom} is reformulated in terms of finite differences, extended to the stiffened gas equation of state (SG EOS) and combined with the LBI factored scheme to stabilize the method around flow discontinuities at high Mach numbers. A novel stabilization term is proposed for low-Mach number applications. The resulting algorithm is shown to be efficient in both low- and high-Mach number regimes. Next, the algorithm is extended to a multicomponent case using an interface capturing strategy (see e.g. \cite{simple}) with surface tension as a continuous surface force (see \cite{sten}). Special care is taken to avoid spurious oscillations of pressure and generation of artificial acoustic waves in the numerical mixture layer. Numerical tests are presented to verify the performance and stability properties for a wide range of flows.

\end{abstract}


\end{frontmatter}

\section{Introduction}
\label{Introduction}

The goal of this study is to develop an algorithm for solving the compressible Navier-Stokes equations (compressible NSEs) usable in various applications, ranging from supersonic shock-interface or shock-boundary layer interactions to nearly incompressible cases, such as highly subsonic astrophysical flows. 
The compressible NSEs have been shown to converge to the incompressible NSEs as the Mach number ($M$) approaches zero (see \cite{klain1981},\cite{klain1982}). However, numerical methods designed for highly compressible flows are known to experience severe problems in the incompressible limit (see e.g. \cite{guil1},\cite{guil2}), such as over-resolution in time due to the dependence of the Courant-Friedrichs-Lewy (CFL) constant on the Mach number, and over-resolution in space due to the incorrect scaling of the artificial viscosity term. 
A number of techniques have been proposed in the literature to overcome these issues, including but not limited to, preconditioners designed to rescale the artificial dissipation term (see e.g. \cite{turkelas},\cite{turkelprec}), and explicit-implicit flux splitting (see e.g. \cite{cordier},\cite{dimarco},\cite{klein}, \cite{note}), constructed to allow for an efficient numerical solution of time-dependent problems. Another way to construct an all-speed algorithm is to develop an efficient way of solving the fully-implicit Euler system/NSEs. Linearized block implicit (LBI) factored schemes, such as algorithms from \cite{brileystruct}, \cite{beamimp}, \cite{beamwarming}, combine a linearization technique based on Taylor expansions with an approximate factorization strategy. 
A related scheme is proposed here for the compressible NSEs based on a factorization strategy from \cite{samvab}, (p. 83). Although it had been reported that implicit algorithms of the ADI-type do not cure the stiffness problem at low Mach numbers (see \cite{turkelprec}), Choi and Merkle demonstrated in \cite{choi} that the reason for the ineffectiveness of these schemes is the splitting error of the factorization, which dominates the physical fluxes as $M \rightarrow 0$. In this study, a reduction strategy, similar to the one used in \cite{maxreduced} for the case of the two-dimensional Maxwell equations, is employed to solve the issue. It allows for the use of the method at extremely low Mach numbers without adding extra computational cost, as is demonstrated in sections \ref{Gresho Vortex.} and \ref{Manufactured Solution.}.
 
Centered-in-space discretization of physical fluxes in a system of hyperbolic conservation laws requires an introduction of artificial dissipation terms to avoid the high-frequency oscillations due to odd-even decoupling at low Mach numbers, and spurious oscillations across flow discontinuities for highly compressible flows. 
Here an artificial dissipation term similar to the one in \cite{invdom} is added implicitly to the scheme at high Mach numbers. A nonlinear adaptive choice of artificial viscosity coefficient based on the maximum wave propagation speed in local one-dimensional Riemann problems (see \cite{maxspeed} for details) guarantees robust behavior of the method for an arbitrary system of hyperbolic conservation laws, while being less dissipative than classical first-order schemes, such as Lax-Friedrichs (see section \ref{Sod Shock Tube.} for details). For the low-Mach number regime this term needs to be rescaled to avoid over-dissipation, similar to other shock-capturing schemes (see \cite{turkelprec}, \cite{mic}, \cite{mic2}). Instead of using a preconditioning matrix to perform such rescaling, or introducing a fourth-order difference term, a different approach is proposed in this paper. A novel artificial dissipation term is designed based on a second-order finite difference with artificial viscosity proportional to $h^2$. This difference operator is applied to the conservative variables and to the product of conservative variables and their corresponding Jacobians. This combination allows one to maintain control of the kinetic energy and dump high-frequency oscillations, and thus to achieve an efficient solution for low-Mach number flows. The effectiveness of the proposed stabilization is shown in Section \ref{Gresho Vortex.}.

Finally, the algorithm allows for the simulation of multicomponent flows with surface tension. Among conventional methods of simulation of material interfaces are interface-capturing (see e.g. \cite{5eq},\cite{wenocol},\cite{sharp},\cite{howabg},\cite{oscsum},\cite{osher},\cite{simple}), interface-tracking (see e.g. \cite{trcher},\cite{jur}), and ghost fluid methods (see e.g. \cite{gfm}). 
The interface-capturing technique based on the advection of the volume of fluid (VoF) function (similar to \cite{wenocol},\cite{oscsum},\cite{simple}) is chosen here alongside the interface sharpening technique from \cite{sharp} for its numerical efficiency, easiness of implementation in higher dimensions, ability to automatically deal with topological changes, and possible implementations of interfacial physics (see e.g. \cite{sten}). Using a diffused interface model and conservative formulation of the governing equations requires special care to avoid spurious pressure oscillations across the interface and generation of artificial acoustic waves in the numerical mixture layer (see e.g. \cite{simple},\cite{5eq}). The consistency of the Equation of State (EOS) in the mixture layer and preservation of contact discontinuity are guaranteed by the appropriate choice of advected flow variables and their incorporation into the computation of Jacobians and the linearization procedure, as well as a special treatment of the VoF-advection equation that is consistent with the rest of the governing equations. These results are in line with the explicit case described in \cite{simple} and provide an extension of the LBI factored schemes to the multicomponent case. The dissipation terms proposed in this paper were also found to preserve the pressure and velocity equilibrium at interfaces.

The paper is organized as follows. A new version of the LBI factored scheme with splitting error reduction is described in Section \ref{Linearized Block ADI Method with Splitting Error Reduction.}. Section \ref{Stabilization.} is dedicated to the artificial dissipation terms at high and low Mach numbers. It finishes the formulation of the algorithm for single-component flows. The algorithm is further extended to the multicomponent case in Section \ref{Interface.}. Summary of the algorithm can be found in Section \ref{Summary of the Method.}. Numerical test cases are presented in Section \ref{Numerical Tests.}. Section \ref{Conclusion.} provides some concluding remarks and discussions on the method and possible directions for future studies.

\section{Linearized Block ADI Method with Splitting Error Reduction.}
\label{Linearized Block ADI Method with Splitting Error Reduction.}
\subsection{Governing Equations and Linearization.}
\label{Governing Equations and Linearization.}
In order to avoid numerical errors introduced by non-conservative schemes in the presence of shock waves (see \cite{hou},\cite{karni}), we consider the compressible Navier-Stokes equations in conservative form.  We demonstrate
the ideas on the 2D version of the equations but the proposed schemes extend easily to three dimensions. For the single-component case, the system can be written as (see \cite{beamwarming}):\\
\begin{equation}
\label{1}
\begin{split}
\frac{\partial U}{ \partial t} + \frac{\partial F(U)}{\partial x} + \frac{\partial G(U)}{\partial y} = 
\frac{\partial V_1 (U,U_x)}{\partial x} &+ \frac{\partial V_2 (U,U_y)}{\partial x} \\+ \frac{\partial W_1(U,U_x)}{\partial y} + \frac{\partial W_2(U,U_y)}{\partial y},
\end{split} 
\end{equation}\\
where $U = (\rho,m,n,E)$ is the vector of conservative variables (density, momentum in x and y direction, and total energy), $F$ and $G$ are the fluxes associated with the Euler system, and $V_1$, $V_2$, $W_1$, $W_2$ are the fluxes associated with the viscous stress tensor (see Appendix A for details).

Using the mixed implicit-explicit Euler time discretization, one can write a semi-discrete version of (\ref{1}) as:\\
\begin{equation}
\label{2}
\begin{split}
\frac{U^{n+1} - U^n}{\tau} + \frac{\partial F^{n+1}(U)}{\partial x} + \frac{\partial G^{n+1}(U)}{\partial y} =  \frac{\partial V_1^{n+1}(U,U_x)}{\partial x} &+ \frac{\partial V_2^{n}(U,U_y)}{\partial x} \\+ \frac{\partial W_1^{n}(U,U_x)}{\partial y} + \frac{\partial W_2^{n+1}(U,U_y)}{\partial y},
\end{split} 
\end{equation} \\
where $\tau$ is the time step. Due to the non-linearity of the fluxes $F$, $G$, $V_1$, and $W_2$, the space-discretization of (\ref{2}) produces a non-linear system of algebraic equations. To approximate the solution of the nonlinear system, a simple linearization, similar to the one from \cite{brileystruct}, can be employed:\\
\begin{equation}
\label{3}
F^{n+1} = F^n + \left ( \frac{\partial F}{\partial U} \right )^n (U^{n+1} - U^n)
\end{equation}
\begin{equation}
\label{4}
G^{n+1} = G^n + \left ( \frac{\partial G}{\partial U} \right )^n (U^{n+1} - U^n)
\end{equation}
\begin{equation}
\label{5}
V_1^{n+1} = V_1^n + \left ( \frac{\partial V_1}{\partial U} \right )^n (U^{n+1} - U^n) + \left ( \frac{\partial V_1}{\partial U_x} \right )^n (U_x^{n+1} - U_x^n)
\end{equation}
\begin{equation}
\label{6}
W_2^{n+1} = W_2^n + \left ( \frac{\partial W_2}{\partial U} \right )^n (U^{n+1} - U^n) + \left ( \frac{\partial W_2}{\partial U_y} \right )^n (U_y^{n+1} - U_y^n).
\end{equation}\\
Substituting expressions (\ref{3})-(\ref{6}) into (\ref{2}) and combining explicit terms (denoted as $R^n$) and implicit terms in corresponding directions (denoted as $\textbf{A}_x U^{n+1}$ and $\textbf{A}_y U^{n+1}$), equation (\ref{2}) can be written as:\\ 
\begin{equation}
\label{7}
\left (  \textbf{I} + \tau \textbf{A}_x + \tau \textbf{A}_y \right ) U^{n+1} = \tau R^n.
\end{equation}

Folowing the approach in \cite{samvab} (p.83), equation (\ref{7}) can be approximated by the following factorized equation:\\
\begin{equation}
\label{11}
\left (  \textbf{I} + \tau \textbf{A}_x)(\textbf{I} + \tau \textbf{A}_y \right ) U^{n+1} = \tau R^n,
\end{equation}\\
with splitting error $\textbf{ER}(U^{n+1}_k) = \tau^2 \textbf{A}_x \textbf{A}_y U^{n+1}$, and solved as a sequence of two one-dimensional problems 
\begin{equation}
\label{12}
(\textbf{I} + \textbf{A}_x) \hat{U}^{n+1} = \tau R^n
\end{equation}
\begin{equation}
\label{13}
(\textbf{I} + \textbf{A}_y) U^{n+1} = \hat{U}^{n+1}.
\end{equation}
Each of these problems requires the solution of block-tridiagonal linear systems only, which can be performed by a block-tridiagonal extension of the Thomas algorithm for tridiagonal systems (e.g. see \cite{fletcher1}, Volume 1, pp.188-189). The parallel implementation of the Thomas algorithm using the Schur complement technique and domain decomposition, as described in \cite{minevadi}, can be easily extended for the block-tridiagonal version of the linear solver. Weak scalability results for the method can be found in Section \ref{Weak Scalability.}.

\begin{remark}
The method can be reformulated for different, more accurate time-marching schemes and different splitting strategies to improve accuracy. Here the implicit Euler method with the splitting from \cite{samvab}, (p.83) are chosen for their simplicity and robustness. In particular, this splitting allows multicomponent factorization of non-commutative operators. Therefore, an extension of the method to the three-dimensional case is possible.
\end{remark}

\subsection{Splitting error reduction.}
\label{Splitting error reduction.}
The splitting error introduced by the factorization (\ref{11}) has been found to grow as the Mach number approaches zero, while in one dimension the LBI factored schemes demonstrate a performance similar to the artificial compressibility method (see \cite{choi} for details). There are several possible ways of reducing the error (see \cite{brileyrefl} for a review). One of the possibilities is to perform the following iterations (here the subscript denotes the iteration level):
\begin{equation}
\label{14}
(\textbf{I} + \textbf{A}_x) \hat{U}^{n+1}_{k+1} = \tau R^n + \textbf{ER}(U^{n+1}_k)
\end{equation}
\begin{equation}
\label{15}
(\textbf{I} + \textbf{A}_y) U^{n+1}_{k+1} = \hat{U}^{n+1}_{k+1},
\end{equation}
 with $U^{n+1}_0 = U^{n}$. This reduction strategy is similar to the one used in \cite{maxreduced} in the context of the two-dimensional Maxwell equations.
 None of the test cases presented in this paper required more than one iteration, i.e. simple addition of $\textbf{ER}(U^{n})$ to the right-hand-side of the system provided sufficient reduction of the splitting error for the tests with low-Mach number flows. Hence, at every time step the following system was solved:
\begin{equation}
(\textbf{I} + \textbf{A}_x) \hat{U}^{n+1} = \tau R^n + \textbf{ER}(U^{n})
\end{equation}
\begin{equation}
(\textbf{I} + \textbf{A}_y) U^{n+1} = \hat{U}^{n+1}.
\end{equation}

Note that the proposed technique is equivalent to the preconditioned Richardson iterative method (see e.g. \cite{rya}, Chapter 6) applied to the linearized system (\ref{7}). Indeed, if
\begin{equation}
\textbf{A} = \textbf{I} + \tau \textbf{A}_x + \tau \textbf{A}_y
\end{equation}
a preconditioner can be defined as:
\begin{equation}
\textbf{P}^{-1} = \left ( \left (  \textbf{I} + \tau \textbf{A}_x)(\textbf{I} + \tau \textbf{A}_y \right) \right)^{-1}.
\end{equation}
Then $\textbf{A} = \textbf{P} - \textbf{ER}$ where $\textbf{ER}= \tau^2 \textbf{A}_x \textbf{A}_y$. Then, the system (\ref{14})-(\ref{15}) is equivalent to
\begin{equation}
\textbf{P} U^{n+1}_{k+1} = \textbf{ER} U^{n+1}_{k} + \tau \textbf{R}^n
\end{equation}
or alternatively
\begin{equation}
U^{n+1}_{k+1} = U^{n+1}_{k} + \textbf{P}^{-1} \textbf{r}_k
\end{equation}
\begin{equation}
\textbf{r}^k = \tau \textbf{R}^n - \textbf{A} U^{n+1}_k.
\end{equation}
With this preconditioner the iterative method converges very rapidly.
\section{Stabilization.}
\label{Stabilization.}
\subsection{Guermond-Popov (GP)  shock-capturing scheme.}
\label{Guermond-Popov shock-capturing scheme.}
In the supersonic regions of the flow, we need to introduce stabilization terms to dissipate the high-frequency oscillations associated with higher-order spatial discretizations.
We use the idea behind the finite element stabilization method for general hyperbolic systems, proposed in \cite{invdom}. It can be reformulated in terms of finite differences as follows:
 \begin{equation}
 \label{16}
 \begin{split}
 \frac{U_{i,j}^{n+1} - U_{i,j}^{n}}{\tau} + \frac{\hat{F}^n_{i+\frac{1}{2},j} - \hat{F}^n_{i-\frac{1}{2},j}}{h_x} + \frac{\hat{G}^n_{i,j+\frac{1}{2}} - \hat{G}^n_{i,j-\frac{1}{2}}}{h_y} = 0
 \end{split}
 \end{equation}
 with numerical numerical fluxes:
 \begin{equation}
 \label{17}
 \hat{F}^n_{i+\frac{1}{2},j} = \frac{F^n_{i+1,j}+F^n_{i,j}}{2} - \lambda_{i+1,j} \left (U^n_{i+1,j} - U^n_{i,j}  \right ).
 \end{equation}
Artificial viscosity coefficients are defined here as:
\begin{equation}
 \lambda_{i+1,j} = \lambda_{max}(U_{i+1,j}^n,U^n_{i,j}),
 \end{equation}
 with $\lambda_{max}(U_l,U_r)$ being a maximum wave speed estimation for the corresponding local one-dimensional Riemann problem with initial conditions given by $U_l$ and $U_r$ (see \cite{invdom},\cite{maxspeed} for details).
 
The original algorithm is fully explicit and preserves all the convex invariant sets, which, in the case of the Euler system, guarantees positivity of density and internal energy, and produces a solution that satisfies the entropy inequality for every entropy pair of the hyperbolic system (see \cite{invdom} for details). These properties are essential for obtaining an approximation of a physical solution to the Euler equations and are achieved without any extra constructions (such as flux limiters or non-oscillatory reconstructions). Unlike Godunov-type schemes, the method does not require an exact or approximate solution of local Riemann problems. Only an estimation of the maximum speed of wave propagation is needed. The efficiency of the algorithm depends on the particular method used to obtain the estimation. The estimation procedure proposed in this study can be found in Section \ref{Maximum Speed estimation for the multicomponent Stiffened Gas EOS.}. The procedure is developed for the case of multiple components obeying Stiffened Gas EOS and includes the effects of surface tension.

\begin{remark}
The original finite element first-order method has been employed to construct a second-order invariant domain preserving approximation -- see \cite{invdom2}. There invariant domain preserving auxiliary solutions obtained from the first-order method are used to define local bounds for a high-order, invariant domain violating but an entropy-consistent algorithm, via a convex limiting process. 
\end{remark}  
 
 \begin{remark}
 If $\lambda_{max}(U_l,U_r) = \frac{h}{4 \tau}$, the finite difference version of the method reproduces the classical Lax-Friedrichs scheme. Due to the nonlinear definition of $\lambda_{max}$ through the maximum wave speed of the local one-dimensional Riemann problems, the GP method achieves the same level of robustness as the Lax-Friedrichs scheme while maintaining much sharper flow discontinuities (see Section \ref{Sod Shock Tube.} for comparison).
 \end{remark}
 
As can easily be seen by substituting expression (\ref{17}) into equation (\ref{16}), the method adds a second-order artificial dissipation term of the type (GP-dissipation term):\\
\begin{equation}
\begin{split}
\textbf{D}^{GP} U = h_x \frac{\partial}{\partial x} \left [ \lambda_{max} \frac{\partial}{\partial x} U^{n+1} \right ] +h_y \frac{\partial}{\partial y} \left [ \lambda_{max} \frac{\partial}{\partial y} U^{n+1} \right ]
\end{split}
\end{equation}
\\
to the original explicit centered-in-space scheme. The same second-order stabilizing term can be added to the implicit scheme for the compressible NSEs in the high-Mach number regime, and this is equivalent to defining the inviscid fluxes as:\\
 \begin{equation}
 \hat{F}^{n+1}_{i+\frac{1}{2},j} = \frac{F^{n+1}_{i+1,j}+F^{n+1}_{i,j}}{2} - \lambda_{i+1,j} \left (U^{n+1}_{i+1,j} - U^{n+1}_{i,j}  \right ),
 \end{equation}\\
and applying the same linearization and factorization procedures after that. This implicit treatment of the dissipation term does not disrupt the block-tridiagonal nature of the resulting linear system and therefore does not increase the cost of the computations.
\subsection{Damping of high-frequency oscillations.}
\label{Damping of high-frequency oscillations.}
Like other artificial dissipation terms designed for stabilization around strong shocks, $\textbf{D}^{GP} U$ starts dominating the physical solution as the Mach number approaches zero. Indeed, in the incompressible limit (where the velocity of the flow is small comparing to the speed of sound, $|u|\ll c$), the maximum wave speed $\lambda = \lambda_{max} = u \pm c \sim c$. The perturbed Euler system in one dimension can be written as:\\
\begin{equation}
\frac{\partial \rho}{\partial t} + \frac{\partial}{\partial x} (\rho u) -h  \frac{\partial}{\partial x} \left ( \lambda 
 \frac{\partial}{\partial x} \rho \right ) = 0
\end{equation}  
\begin{equation}
\frac{\partial \rho u}{\partial t} +  \frac{\partial}{\partial x} \left ( \rho u^2 + p \right ) -h   \frac{\partial}{\partial x} \left ( \lambda 
 \frac{\partial}{\partial x} (\rho u)  \right )= 0
\end{equation}
\begin{equation}
\frac{\partial E}{\partial t} +  \frac{\partial}{\partial x} ((E+p)u) -h  \frac{\partial}{\partial x}  \left ( \lambda 
 \frac{\partial}{\partial x} E  \right ) = 0.
\end{equation}\\
Then, introducing the characteristic scales: length $L$, density $\tilde{\rho}$, and velocity $\tilde{U}$, dimensionless variables can be defined as $x^* = \frac{x}{L}$, $\rho^* = \frac{\rho}{\tilde{\rho}}$, $u^* = \frac{u}{\tilde{U}}$, $t^* = \frac{t}{\tilde{U}/L}$, $M = \frac{\tilde{U}}{c}$, $p^* = \frac{p}{\tilde{\rho} c^2}$, and $E^* = \frac{E}{\tilde{\rho} c^2}$.\\
Since $\lambda \sim c$, the system can be rewritten in the following non-dimensional form:\\
\begin{equation}
\frac{\partial \rho^*}{\partial t^*} + \frac{\partial}{\partial x^*} (\rho^* u^*) -\frac{h}{M}  \frac{\partial^2}{\left ( \partial x^* \right )^2} \rho^*= 0
\end{equation}  
\begin{equation}
\frac{\partial \rho^* u^*}{\partial t^*} +  \frac{\partial}{\partial x^*} \left ( \rho^* \left ( u^* \right )^2 + \frac{p^*}{M^2} \right ) -\frac{h}{M}  \frac{\partial^2}{\left ( \partial x^* \right )^2} (\rho^* u^*)= 0
\end{equation}
\begin{equation}
\frac{\partial E^*}{\partial t^*} +  \frac{\partial}{\partial x^*} ((E^*+p^*)u^*) -\frac{h}{M} 
 \frac{\partial^2}{\left ( \partial x^* \right )^2} E^*= 0.
\end{equation}\\
Hence, the artificial dissipation term indeed becomes dominant as $M \rightarrow 0$ if $h$ is fixed.

At low Mach numbers, the following dissipation term is proposed in this paper (Low-Mach (LM-) dissipation term):
\begin{equation}
\begin{split}
\textbf{D}^{LM} & U^{n+1} = \textbf{D}^{LM}_1 U^{n+1} + \textbf{D}^{LM}_2 U^{n+1} =\\ h^2 \omega_1 \nabla^2 U^{n+1} + h^2 \omega_2 & \left (\frac{\partial^2}{\partial x^2} \left [ \left (\frac{\partial F}{\partial U} \right )^n U^{n+1} \right ] + \frac{\partial^2}{\partial y^2} \left [ \left (\frac{\partial G}{\partial U} \right )^n U^{n+1} \right ] \right ),  
\end{split}
\end{equation}
where $\omega_1$ and $\omega_2$ are scalar dimensionless parameters.

The purpose of the first term here (which is the same as the GP-term with $\lambda \equiv h$) is to drain off the kinetic energy and thus maintain the overall stability of computations. The second and the third terms are used to regularize the entries of the corresponding matrix operator and connect odd and even nodes, which are decoupled due to the use of the centered-in-space discretization for hyperbolic fluxes. Jacobians provide proper weights for every variable depending on the magnitude of its contribution to the linearized flux. The effects of both of these terms at different values of the Mach number are illustrated by numerical examples in Section \ref{Gresho Vortex.}.

\begin{remark}
A related idea of matching the scaling of the artificial viscosity matrix with the scaling of the Jacobians in order to stabilize the system in the low-Mach number limit was used in \cite{mic} in the context of the Roe-Turkel scheme. However, here instead of choosing scaling parameters to match orders of magnitude of the Jacobian's components, these components themselves are used in the stabilizing term. This was found to be more efficient in the context of the present scheme. Furthermore, as is revealed in the Section \ref{Preservation of contact discontinuities.}, the artificial dissipation term does not disrupt the velocity and pressure equilibrium at interfaces and is therefore compatible with the Volume of Fluid method.
\end{remark}

As for the GP-dissipation term, the LM-term can be incorporated into the factorization strategy in a straightforward way with direction splitting. Since the splitting error due to the LM-term involves the components of the Jacobian, it should be taken into account while performing the reduction strategy described above.

If the flow under consideration features regions of low and high Mach numbers, switching between LM- and GP-dissipation terms can be performed based on the local Mach number $M_{loc}$ as follows:
\begin{equation}
\label{24}
\textbf{D} U^{n+1} = k^{(HM)} \textbf{D}^{GP} U^n + k^{(LM)}\textbf{D}^{LM} U^n,
\end{equation}
where $k^{(HM)}$,  $k^{(LM)}$ are scalar dimensionless parameters to be defined.
Using an approximation to the Heaviside function and setting a threshold Mach number $M_{tr}$, $k^{(HM)}$ and $k^{(LM)}$ are defined as:
\begin{equation}
k^{(HM)} = \frac{1}{1+ e^{-2k(M_{loc} - M_{tr})}},
\end{equation}
\begin{equation}
k^{(LM)} = 1 - k^{(HM)}.
\end{equation}
The described strategy has been shown to be efficient for Mach numbers as low as $M=10^{-6}$. See Section \ref{Gresho Vortex.} for details and numerical illustrations.

\begin{remark}
The parameters $\omega_1$ and $\omega_2$ still have to be tuned manually depending on the particular application. A rigorous theoretically justified adaptive scaling of these coefficients can lead to an improvement of the algorithm and will be considered in the future. 
\end{remark}
The operators $\textbf{A}_x$ and $\textbf{A}_y$ are then redefined to include the implicit artificial dissipation terms in the corresponding directions.
\section{Interface Capturing and Computation of the Surface Tension.}
\label{Interface.}
The interface capturing is based on the following advection equation for the volume of fluid (VoF) function $\phi$ (see e.g. \cite{5eq},\cite{simple},\cite{howabg},\cite{oscsum},\cite{osher},\cite{wenocol}):
\begin{equation}
\label{27}
	\frac{\partial \phi}{\partial t} + \vec{u} \cdot \nabla \phi = 0,
\end{equation}
where $\vec{u} = (\frac{m}{\rho},\frac{n}{\rho}) = (u,v)$ is the flow velocity.
Initially, $\phi$ is set to $1$ for the first fluid and $0$ for the second one. Due to numerical diffusion, this initial discontinuity is smoothed out and leads to the formation of an artificial mixing layer where $\phi \in (0,1)$. EOS parameters have to be consistently defined in the layer to avoid spurious pressure oscillations that appear when a conservative formulation is employed (see e.g. \cite{5eq},\cite{osher},\cite{oscsum},\cite{karni}). Here we use the Stiffened Gas EOS (SG EOS) (see \cite{sgeos}):
\begin{equation}
\label{28}
p = (\gamma - 1)\rho e - \gamma \pi_{\infty},
\end{equation}
where $\gamma$ and $\pi_{\infty}$ are constants for a given fluid, $e$ is the internal energy, defined implicitly through
\begin{equation}
E = \rho e + \rho \frac{u^2 + v^2}{2}.
\end{equation}
The SG EOS is commonly used to model compressible multicomponent flows of gases, liquids, and solids (e.g. air, water, copper, uranium) (see \cite{wenocol} for details). Note that the ideal gas EOS is a particular case of the SG EOS with $\pi_{\infty} = 0$.
\subsection{Barotropic Mixture Laws.}
\label{Barotropic Mixture Laws.}
If $\rho_i$, $e_i$, $p_i$ ($i=1,2$) are density, internal energy and pressure of corresponding fluids, then:
\begin{equation}
\rho = \phi \rho_1 + (1-\phi) \rho_2
\end{equation}
\begin{equation}
\rho e = \phi \rho_1 e_1 + (1-\phi) \rho_2 e_2
\end{equation}
\begin{equation}
p = \phi p_1 + (1-\phi) p_2.
\end{equation}
In the regions of pure fluid 1 or pure fluid 2, the pressure is given by the EOS (\ref{28}). In the artificial mixing layer, where both components are present, the following condition must be satisfied (see \cite{5eq}):
\begin{equation}
p_1 = p_2 = p.
\end{equation}
Then
\begin{equation}
\begin{split}
\rho e =& \phi \frac{p + \gamma_1 \pi_1}{\gamma_1 - 1} + (1-\phi)\frac{p + \gamma_2 \pi_2}{\gamma_2 - 1} = \\
p (\frac{\phi}{\gamma_1 - 1} &+ \frac{1-\phi}{\gamma_2-1}) + (\frac{\phi \gamma_1 \pi_1}{\gamma_1 - 1} + \frac{(1-\phi) \gamma_2 \pi_2}{\gamma_2-1}),
\end{split}
\end{equation}
which leads to the following definition of the proper averaging for the EOS coefficients in the mixing layer ($\gamma$, $\pi^{\infty}$):
\begin{equation}
\label{34}
\alpha = \frac{1}{1-\gamma} = \phi \frac{1}{1-\gamma_1} + (1-\phi)\frac{1}{1-\gamma_2}
\end{equation}
\begin{equation}
\label{35}
\beta = \frac{\pi^{\infty}\gamma}{1-\gamma} = \phi \frac{\pi^{\infty}_1\gamma_1}{1-\gamma_1} + (1-\phi)\frac{\pi^{\infty}_2\gamma_2}{1-\gamma_2}.
\end{equation}
Further considerations will reveal the important role of the parameters $\alpha$ and $\beta$ in preserving contact discontinuities at interfaces.

\subsection{Preservation of contact discontinuities at interfaces.}
\label{Preservation of contact discontinuities.}
Another requirement needed to preserve the pressure and velocity equilibrium in multicomponent flows is that if the velocity and pressure are constants at the interface at time $t^n$ they remain constants at $t^{n+1}$ (preservation of a contact discontinuity at interfaces) (see \cite{simple} for details):

\begin{center}
if $u_i^{n} = u = const$ and $p_i^{n} = p = const$ for any $i$, then $u_i^{n+1} = u$ and $p_i^{n+1} = p$.
\end{center}

To avoid technicalities, the one-dimensional inviscid case will be considered here. The following analysis can be easily extended to the multidimensional case. The viscous part of the equations does not disrupt the velocity and pressure equilibrium since it only involves terms with velocity derivatives.

For a one-dimensional inviscid flow, system (\ref{11}) becomes (see Appendix A for details):
\begin{equation}
\label{C1}
\begin{split}
U^{n+1} + \tau \frac{\partial}{\partial x} \left [ \left ( \frac{\partial F}{\partial U}  \right )^n U^{n+1} \right ] = U^n + \tau \frac{\partial}{\partial x} \left [ \left ( \frac{\partial F}{\partial U}  \right )^n U^n - F^n  \right ] \\+ \tau D_1(U^{n+1}) + \tau D_2(U^{n+1}) + \tau D_3(U^{n+1}),
\end{split}
\end{equation}
where\\
\begin{equation}
D_1(U^{n+1}) = h \frac{\partial}{\partial x} \left [ \lambda \frac{\partial}{\partial x} U^{n+1} \right ]
\end{equation}
\begin{equation}
D_2(U^{n+1}) = h^2 \frac{\partial^2}{\partial x^2}  U^{n+1}
\end{equation}
\begin{equation}
D_3(U^{n+1}) = h^2 \frac{\partial^2}{\partial x^2} \left [ \left ( \frac{\partial F}{\partial U} \right )^n U^{n+1} \right ].
\end{equation}\\
Here, $\alpha$ and $\beta$ are included in the set of variables, thus extending the set $U = (\rho,m, E)$ to $U = (\rho,m, E,\alpha,\beta) = (\rho,\rho u, \frac{\rho u^2}{2} + \beta + \alpha p,\alpha,\beta)$). The advection equation for the interface is written for $\alpha$ and $\beta$, rather than for the VoF function $\phi$. The following analysis reveals that this is the natural choice of variables to guarantee the desired preservation property.

Given that $u_i^{n} = u$ and $p_i^{n} = p$, it is assumed that the same holds at the next time level, i.e. $u_i^{n+1} = u$ and $p_i^{n+1} = p$. Then, we check if this assumption provides the solution for the discrete linearized mass, momentum and energy equations. 

Using Appendix A:
\begin{equation}
F^n = 
\begin{bmatrix}
\rho^n u & \rho^n u^2 + p & \frac{\rho^n u^3}{2} + u \beta^n + u p \alpha^n + u p 
\end{bmatrix}
^T
\end{equation}

\begin{equation}
\left ( \frac{\partial F}{\partial U} \right )^n = \\
\begin{bmatrix}
0 & 1 & 0 & 0 & 0 \\
- u^2 + \frac{u^2}{2 \alpha^n}& 2 u - \frac{u}{\alpha^n}& \frac{1}{\alpha^n}& \frac{p}{\alpha^n}& -\frac{1}{\alpha^n} \\
\delta_1 & \delta_2 & u \left ( 1 + \frac{1}{\alpha^n}  \right ) & -\frac{up}{\alpha^n} & -\frac{u}{\alpha^n}
\end{bmatrix}
\end{equation}
where  $\delta_1 = -\frac{\left ( \alpha^n p + p + \beta^n + \frac{\rho^n u^2}{2}  \right )u}{\rho^n} + \frac{u^3}{2 \alpha^n}$, $\delta_2 = \frac{\alpha^n p + p + \beta^n + \frac{\rho^n u^2}{2}}{\rho^n} - \frac{u^2}{\alpha^n}$.

\begin{equation}
\label{C4}
\left ( \frac{\partial F}{\partial U} \right )^n  U^n = 
\begin{bmatrix}
u \rho^n & u^2 \rho^n & \frac{u^3 \rho^n}{2} + u \beta^n + u p \alpha^n
\end{bmatrix}
^T
\end{equation}

\begin{equation}
\left ( \frac{\partial F}{\partial U} \right )^n  U^{n+1} = 
\begin{bmatrix}
u \rho^{n+1} & u^2 \rho^{n+1} & \frac{u^3 \rho^{n+1}}{2} + u \beta^{n+1} + u p \alpha^{n+1}
\end{bmatrix}
^T.
\end{equation}
Hence,
\begin{equation}
\left ( \frac{\partial F}{\partial U} \right )^n  U^n - F^n = 
\begin{bmatrix}
0 & -p & - u p
\end{bmatrix}
^T
\end{equation}
 and 
 \begin{equation}
\frac{\partial}{\partial x} \left [ \left ( \frac{\partial F}{\partial U} \right )^n  U^n - F^n \right ] = 
\begin{bmatrix}
0 & 0 & 0
\end{bmatrix}
^T.
\end{equation}\\
Using (\ref{C4}), the mass conservation equation in a discrete form reads as:
\begin{equation}
\label{C8}
\rho^{n+1}_i + \tau u \frac{\rho^{n+1}_{i+1} - \rho^{n+1}_{i-1}}{h} - \rho^n_i - D_1(\rho^{n+1}) - D_2(\rho^{n+1}) -uD_3(\rho^{n+1})= 0,
\end{equation}
and the momentum conservation as:
\begin{equation}
\begin{split}
\rho^{n+1}_i u + \tau u^2 \frac{\rho^{n+1}_{i+1} - \rho^{n+1}_{i-1}}{h} - \rho^n_iu - D_1(\rho^{n+1} u) - D_2(\rho^{n+1} u) \\-u D_3(\rho^{n+1} u) = \\ u \left ( \vphantom{\frac{\rho^{n+1}_{i+1}-\rho^{n+1}_{i-1}}{h}} \rho^{n+1}_i + \tau u \frac{\rho^{n+1}_{i+1}-\rho^{n+1}_{i-1}}{h} - \rho^n_i   - D_1(\rho^{n+1})  - D_2(\rho^{n+1}) \right. \\ \left. -u D_3(\rho^{n+1}) \vphantom{\frac{\rho^{n+1}_{i+1}-\rho^{n+1}_{i-1}}{h}} \right ) = 0,
\end{split}
\end{equation}
which is satisfied if (\ref{C8}) is. Hence, the assumed solution satisfies the mass and momentum parts of (\ref{C1}). It was achieved by including $\alpha$ and $\beta$ into the set of variables.

The energy equation can be written as:
\begin{equation}
\begin{split}
E^{n+1} + \tau u \left ( \frac{\rho^{n+1}_{i+1}u^2 - \rho^{n+1}_{i-1}u^2}{2 h} + \frac{\beta^{n+1}_{i+1} - \beta^{n+1}_{i-1}}{h} + p\frac{\alpha^{n+1}_{i+1} - \alpha^{n+1}_{i-1}}{h}   \right )\\ - E^n - D_1(E^{n+1})- D_2(E^{n+1})- D_3(E^{n+1}) = 0.
\end{split}
\end{equation} 
Expressing energy through the EOS as $E = \alpha p + \frac{u^2}{2}\rho + \beta$, and substituting it into the previous equations allows one to rewrite it as:\\
\begin{equation}
\label{C11}
\begin{split}
\frac{u^2}{2} \left ( \rho_i^{n+1} + \tau u \frac{\rho^{n+1}_{i+1} - \rho^{n+1}_{i-1}}{h} - \rho^n - D_1(\rho^{n+1}) - D_2(\rho^{n+1}) - \right. \\ \left. \vphantom{\tau u \frac{\rho^{n+1}_{i+1} - \rho^{n+1}_{i-1}}{h}} uD_3(\rho^{n+1}) \right ) +\\ p \left (  \alpha_i^{n+1} + \tau u \frac{\alpha^{n+1}_{i+1} - \alpha^{n+1}_{i-1}}{h} - \alpha^n- D_1(\alpha^{n+1}) - D_2(\alpha^{n+1})  \right. \\ \left. \vphantom{\frac{\alpha^{n+1}_{i+1} - \alpha^{n+1}_{i-1}}{h}} - uD_3(\alpha^{n+1})   \right ) \\+ \left (  \beta_i^{n+1} + \tau u \frac{\beta^{n+1}_{i+1} - \beta^{n+1}_{i-1}}{h} - \beta^n  - D_1(\beta^{n+1}) - D_2(\beta^{n+1})  \right. \\ \left. \vphantom{\frac{\beta^{n+1}_{i+1} - \beta^{n+1}_{i-1}}{h}} -u D_3(\beta^{n+1}) \right ) = 0.
\end{split}
\end{equation}\\
The expression in the first brackets in equation (\ref{C11}) is zero if the mass equation (\ref{C8}) is satisfied. The expressions in the second and third brackets 
are equal to zero if the discretization of the advection equations for $\alpha$ and $\beta$:
\begin{equation}
\frac{\partial \alpha}{\partial t} + \vec{u} \cdot \nabla \alpha = 0
\end{equation}
and
\begin{equation}
\frac{\partial \beta}{\partial t} + \vec{u} \cdot \nabla \beta = 0,
\end{equation}
is the same as for the energy equation:
\begin{equation}
\begin{split}
\frac{\alpha^{n+1}_i - \alpha^n_i}{\tau} + u^n \frac{\alpha^{n+1}_{i+1} - \alpha^{n+1}_{i-1}}{h} - D_1(\alpha^{n+1}) - D_2(\alpha^{n+1}) \\- D_3(u^n\alpha^{n+1})= 0
\end{split}
\end{equation}
\begin{equation}
\begin{split}
\frac{\beta^{n+1}_i - \beta^n_i}{\tau} + u^n \frac{\beta^{n+1}_{i+1} - \beta^{n+1}_{i-1}}{h} - D_1(\beta^{n+1}) - D_2(\beta^{n+1}) \\- D_3(u^n\beta^{n+1})= 0.
\end{split}
\end{equation}
This analysis can be trivially extended to the multidimensional case, where the same argument leads to the use of the same factorization and splitting error reduction strategies for the interfacial advection as for the rest of the system.

Hence, similar to the Godunov-Rusanov scheme from \cite{simple}, the contact discontinuity preservation condition determines the discretization of the advection equation for the VoF function. Furthermore, this analysis reveals that the parameters $\alpha$ and $\beta$ are the natural choice for advected variables and should be taken into account when constructing the Jacobians and performing the linearization. 

However, solving two similar advection equations for both $\alpha$ and $\beta$ can be avoided. Indeed, they both follow from the solution of (\ref{27}) and (\ref{34})-(\ref{35}). $\phi^{n+1}$ can be found before updating other variables and used to compute $\alpha^{n+1}$ and $\beta^{n+1}$. This allows one to compute their contributions to the remaining system as a part of the right-hand-sides of (\ref{12})-(\ref{13}).

In the most general case, the VoF function is updated through an iterative (until the overall splitting error is reduced enough) solution of:
\begin{equation}
\label{vf1}
(\textbf{I} + \tau u^n \partial_x - \tau \textbf{D}_x) \hat{\phi}^{n+1}_{k+1} = \phi^n + \textbf{ER}(\phi^{n+1}_k)
\end{equation}
\begin{equation}
\label{vf2}
(\textbf{I} + \tau v^n \partial_y - \tau \textbf{D}_y) \phi^{n+1}_{k+1} = \hat{\phi}^{n+1}
\end{equation}
where 
\begin{equation}
\textbf{ER}(\phi^{n+1}_k) = \tau^2 ( u^n \partial_x -  \textbf{D}_x) (v^n \partial_y - \textbf{D}_y).
\end{equation}

\subsection{Interface sharpening.}
\label{Interface sharpening.}
Interface-capturing schemes are known to diffuse interfaces during the course of computations, due to the introduction of artificial dissipation. A sharpening algorithm is proposed in \cite{sharp} to keep the width of the interface constant (typically several grid cells). To achieve this, $\rho$ and $\phi$ are updated by solving the Euler explicit discretization of the following artificial compression equations until a steady state with some predefined tolerance $tol$:
\begin{equation}
\frac{\partial \rho}{\partial T} = H(\phi) \vec{n} \cdot \left ( \nabla \left (\epsilon_h \vec{n} \cdot \nabla \rho \right ) - (1-2 \phi) \nabla \rho \right )
\end{equation}
\begin{equation}
\frac{\partial \phi}{\partial T} = \vec{n} \cdot (\epsilon_h |\nabla \phi| - \phi(1 - \phi)),
\end{equation}
where $T$ is an artificial time-like parameter, $\epsilon_h$ is the parameter that controls the thickness of the interface, and
\begin{equation}
H(\phi) = tanh((\frac{\phi(1-\phi)}{10^{-2}})^2),
\end{equation}
is a regularized $\delta$-function that limits the artificial compression to the interfacial layer and prevents the density update to influence other types of discontinuities, such as shock waves.

The parameter $\epsilon_h$ is typically chosen to be proportional to the grid size, as $\epsilon_h = \epsilon \cdot h$, where $\epsilon$ is a scalar dimensionless parameter to be defined for each particular application. The algorithm has been shown to be robust and efficient, even for high density ratios.

However, this algorithm also maintains a relatively sharp interfacial profile that hampers the computation of the interface normal. Therefore, \cite{sharp} proposed to use the following function:\\
\begin{equation}
\psi = \frac{\phi^{\alpha}}{\phi^{\alpha}+(1-\phi)^{\alpha}}, \alpha < 1
\end{equation}\\
for computation of the normal vector, as:\\
\begin{equation}
\label{41}
\textbf{n} = \frac{\nabla \phi}{|\nabla \phi|} = \frac{\nabla \psi}{|\nabla \psi|}.
\end{equation}\\

\subsection{Surface tension effects.}
\label{Surface tension effects.}
One of the advantages of the interface-capturing approach is a relatively straightforward implementation of surface tension interfacial effects, which can be incorporated into the model by adding a continuous surface force (CSF) explicitly to the governing equation. Reference \cite{sten} proposes two possible approaches, non-conservative and conservative formulations of the CSF.

The first approach adds the following contributions to the momentum and energy equations (assuming that $\phi = 1$ for the liquid phase, $\phi = 0$ for the gas phase):
\begin{equation}
F^{ST}_{m} = \sigma \kappa \nabla \phi
\end{equation}
\begin{equation}
F^{ST}_{E} = \sigma \kappa \vec{u} \cdot \nabla \phi,
\end{equation}
where $\sigma$ $[\frac{N}{m}]$ is the surface tension coefficient, $\kappa = - \nabla \cdot \frac{\nabla \phi}{|\nabla \phi|}$ is the interfacial curvature.

An alternative way is to redefine the total energy by including a term associated with the interfacial energy:
\begin{equation}
\hat{E} = E + \sigma |\nabla \phi |
\end{equation}
and write the CSF as:\\
\begin{equation}
\label{45}
F^{ST}_{m} = - \nabla \cdot \left ( - \sigma \left  ( |\nabla \phi|\textbf{I} - \frac{\nabla \phi \otimes \nabla \phi}{|\nabla \phi|} \right )  \right )
\end{equation}
\begin{equation}
\label{46}
F^{ST}_{E} = - \nabla \cdot \left ( - \sigma \left  ( |\nabla \phi|\textbf{I} - \frac{\nabla \phi \otimes \nabla \phi}{|\nabla \phi|} \right ) \cdot \vec{u} \right ).
\end{equation}\\

The authors of \cite{sten} claim that the conservative approach (i.e. equations (\ref{45}), and (\ref{46})) may lead to an attenuation of parasitic currents (see \cite{stenergy},\cite{jametst}), however other researchers (see \cite{st2}) associate the parasitic currents with curvature computations and related errors, rather than a surface tension model. A comparison of different CSF models and curvature approximations in the context of the present numerical method is outside of the scope of this paper. 

In this study, the non-conservative approach is used, and the interfacial curvature is computed based on the interfacial normal defined by (\ref{41}) and the centered-in-space discretization.

\begin{remark}
The present method allows a simple extension: the CSF can be added implicitly (in the conservative or non-conservative formulation) as another flux and linearized in a similar way as all the other fluxes in (\ref{2}). It may be expected that such implicitness in the computation of CSF could help to relax the CFL conditions associated with capillary effects, which is of the type $\tau \leq const \cdot \left (\frac{h^3}{\sigma} \right )^{\frac{1}{2}}$ (see \cite{st2}).
\end{remark}
\subsection{Maximum speed estimation for the multicomponent Stiffened Gas EOS.}
\label{Maximum Speed estimation for the multicomponent Stiffened Gas EOS.}
The Guermond-Popov shock-capturing algorithm requires an estimation from above of the maximum speed of propagation in local one-dimensional Riemann problems. The procedure proposed in \cite{maxspeed} has to be modified for the multicomponent case when both fluids obey the SG EOS, and the interfacial jump in pressure is present due to the interfacial curvature and surface tension. Suppose the following parameters are given:
($\rho_l$, $u_l$, $e_l$, $p_l$, $\gamma_l$, $\pi^{\infty}_l$,$\phi_l$) and ($\rho_r$, $u_r$, $e_r$, $p_r$, $\gamma_r$, $\pi^{\infty}_r$,$\phi_r$),
and each of them obeys the EOS:
\begin{equation} 
p_z = (\gamma_z - 1)\rho_z e_z - \gamma_z \pi^{\infty}_z
\end{equation}
where $z=l, r$.
The speed of sound can be computed as:
\begin{equation}
c_z = \sqrt{\frac{\gamma_z (p_z + \pi_z^{\infty})}{\rho_z}}.
\end{equation}

The following one-dimensional Riemann problem is considered (see \cite{maxspeed} for details):
\begin{equation}
\partial_t U + \partial_x (F(U) \cdot \textbf{n})) = 0,
\end{equation}
where $\textbf{n}$ is a unit vector normal to a face of a finite difference cell, with the piecewise constant initial conditions described above. The problem is strictly hyperbolic and its solution consists of two genuinely nonlinear waves (shock or rarefaction) and one linearly degenerate middle wave (contact discontinuity) connecting left and right initial states (see \cite{toro}, Chapter 4 for details). The VoF function $\phi$ is assumed to have zero jumps across shock and rarefaction waves, and contact discontinuities. If $\lambda_1^-$ and $\lambda_3^+$ are two extreme wave speeds, a maximum speed of wave propagation is defined as: 
\begin{equation}
\lambda_{max} = max((\lambda_1^-)_-,(\lambda_3^+)_+),
\end{equation}
where $(\lambda_1^-)_- = max(0,-\lambda_1^-)$, $(\lambda_3^+)^+ = max(0,\lambda_3^+)$. This is the quantity used in the GP dissipation term.

Denoting intermediate pressures by $p_l^*$ and $p_r^*$, the following interfacial condition must be satisfied:
\begin{equation}
\label{A5}
      p_r^* = p_l^* + \kappa \sigma,
\end{equation}
where $\kappa = - \nabla \cdot \frac{\nabla \phi}{|\nabla \phi|}$ is the interfacial curvature. The case of $\sigma = 0$ is considered first, then an extension of the proposed solution for nonzero values of $\sigma$ is discussed. 

Similar to the ideal gas case, described in \cite{toro}, Chapter 4, the function 
\begin{equation}
\label{A6}
\eta (p) = f_l(p) + f_r(p) + u_r - u_l
\end{equation}
is considered, where $f_z$ is a shock curve if $p \geq p_z$ 
\begin{equation}
f_z(p) = (p-p_z) \sqrt{\frac{2}{(\gamma_z+1)\rho_z}} \left (p + \frac{\gamma_z-1}{\gamma_z+1} p_z + \frac{\gamma_z}{\gamma_z+1} \pi^{\infty}_z \right )^{-\frac{1}{2}}
\end{equation}
and a rarefaction curve otherwise (see \cite{sgeosrie})
\begin{equation}
f_z(p) = \frac{2 c_z}{\gamma-1} \left ( \left ( \frac{p+\pi^{\infty}_z}{p_z+\pi^{\infty}_z}  \right )^{\frac{\gamma_z - 1}{2 \gamma_z}} -1 \right )
\end{equation}
with $z=l, r$.

An intermediate pressure $p_l^* = p_r^* = p^*$ is given then as a solution of (see \cite{toro}, Chapter 4):
\begin{equation}
\label{A9}
\eta(p^*) = 0.
\end{equation}
If it is solved for $p^*$, then 
\begin{equation}
\lambda_1^- = u_l -c_l, \text{  }\lambda_3^+ = u_r +c_r
\end{equation}
for rarefaction waves, and
\begin{equation}
\lambda_1^- = u_l -\frac{Q_l}{\rho_l}, \text{  }\lambda_3^+ = u_r+\frac{Q_r}{\rho_r}
\end{equation}
for shocks, where $Q_z$ are the corresponding mass fluxes,
\begin{equation}
Q_z = \sqrt{\frac{(p^* + \frac{\gamma-1}{\gamma+1}p_z + \frac{\gamma}{\gamma+1}\pi^{\infty}_z)(\gamma_z+1)\rho_z}{2}}.
\end{equation}
Finally, 
\begin{equation}
\label{A15}
\lambda_{max} = max((\lambda_1^-)_-,(\lambda_r^+)_+).
\end{equation}
Now, since both $f_l$ and $f_r$ are monotonically increasing and concave down, the Newton-Secant method can be used to solve (\ref{A9}) for $p^*$. Given $p_n^1 < p^* < p_n^2$  one can compute
\begin{equation}
p^1_{n+1} = p_n^1 - \frac{\eta (p_n^1)}{\eta ' (p_n^1)}
\end{equation}
\begin{equation}
p^2_{n+1} = p_n^2 - \eta (p^2_n) \frac{p^2_n- p^1_n}{\eta (p^2_n) - \eta (p^1_n)}
\end{equation}
until convergence. The procedure gives the estimation of $p^*$ from above and below, which is used to estimate the maximum speed of propagation.

Shocks and rarefaction waves can be distinguished by using the signs of $\eta(p_l)$ and $\eta(p_r)$ (positive for rarefaction waves, negative for shock waves). If the corresponding wave is a rarefaction wave, $p^*$ is not needed for the speed estimation and the iterative process can be skipped by immediately computing the corresponding wave speed.

Assuming now a non-zero $\sigma$, (\ref{A6}) can be reformulated as:
\begin{equation}
f_l(p_l^*) + f(p_r^*) + u_r - u_l = 0 
\end{equation}
which needs to be solved in terms of one of the pressures, then the other one can be computed using (\ref{A5}) if needed (i.e. if the corresponding wave is a shock).
\section{Summary of the Method.} \label{Summary of the Method.}
The overall algorithm can be summarized in the pseudocode format as following:\\~\\
$t = 0$\\
Define initial conditions for $[\rho,m,n,E,\phi]$\\
Compute $u^0$,$v^0$,$p^0$\\
Initialize $\phi^0$ and $\rho^0$ using the interface sharpening procedure (Section \ref{Interface sharpening.})\\
Initialize $U^0 = [\rho^0,m^0,n^0,E^0,\alpha^0,\beta^0]$ using  $\phi^0$,$\rho^0$,$u^0$,$v^0$\\
$n = 0$\\
\textbf{DO WHILE} $t<T_{end}$\\
\tab $t=t+\tau$\\
\tab Compute all the Jacobian matrices using $U^n$ (see Appendix A)\\
\tab Compute all the artificial viscosity coefficients for the GP-dissipation term\\
\tab \textbf{IF} (reduction (Section \ref{Splitting error reduction.}) is on) \textbf{THEN}\\
\tab \tab $U^{n+1}_{0} = U^{n}$, $\phi^{n+1}_{0} = \phi^{n}$\\
\tab \textbf{ELSE}\\
\tab \tab $U^{n+1}_{0} = 0$, $\phi^{n+1}_{0} = 0$\\
\tab \textbf{END IF}\\
\tab k = 0\\
\tab \textbf{DO WHILE} ($ER(U^{n+1}_{k})>tolerance$) \\
\tab \tab Solve advection of the VoF function ((\ref{vf1}) and (\ref{vf2})) for $\phi^{n+1}_{k+1}$\\
\tab \tab Solve (\ref{14}) and (\ref{15}) for $U^{n+1}_{k+1}$\\
\tab \tab Compute $ER(U^{n+1}_{k+1})$\\
\tab \tab k = k+1\\
\tab \textbf{END DO}\\
\tab Compute $u^{n+1}$, $v^{n+1}$, and $p^{n+1}$\\
\tab Use interface sharpening (Section \ref{Interface sharpening.}) to update $\rho^{n+1}$ and $\phi^{n+1}$\\
\tab Update $U^{n+1}$ using new values of $\rho^{n+1}$, $\phi^{n+1}$, $u^{n+1}$, $v^n+1$, and $p^{n+1}$\\
\tab $n=n+1$\\ 
\textbf{END DO}
\section{Numerical Tests.}
\label{Numerical Tests.}
In this section, numerical tests are presented to verify the performance of the proposed method. Note that no analytical or extensive numerical study on the stability region has been attempted. For each test, time step ($\tau$) and space step ($h$) are chosen to guarantee the desired accuracy and stability, but the choice may not be optimal. Based on the performed tests, low-Mach number applications do not impose significant stability restrictions, i.e. they allow the value of $\tau$ to be at least several times higher than $h$. Stability regions in high-Mach number applications depend on the types and severity of discontinuities in the flow. The stability regions for our method were found to be larger than ones for the explicit version of the same method, for which CFL-condition is of the type $\tau < const \cdot \frac{h}{|\lambda_{max}|}$ where $|\lambda_{max}|$ is the maximum speed of propagation (see \cite{invdom} for stability analysis of the explicit finite element Guermond-Popov scheme). The use of explicit non-reflection boundary conditions, as in Sections \ref{Shock Wave Refraction.} and \ref{Shock Wave-Bubble Interaction.}, is expected to strengthen the stability restrictions (e.g. see \cite{chak}). 
\subsection{Gresho Vortex.}
\label{Gresho Vortex.}
The behavior of the scheme in the low-Mach number limit is investigated using the Gresho vortex test (similar to \cite{mic2}), a time-independent inviscid rotating vortex placed in a square periodic domain. Scaling the pressure with respect to the reference Mach number allows the study of stability and dissipative properties of numerical methods for different Mach numbers (see \cite{mic} as an example).  

\begin{figure}
 \centering
		    \includegraphics[width=10.5cm]{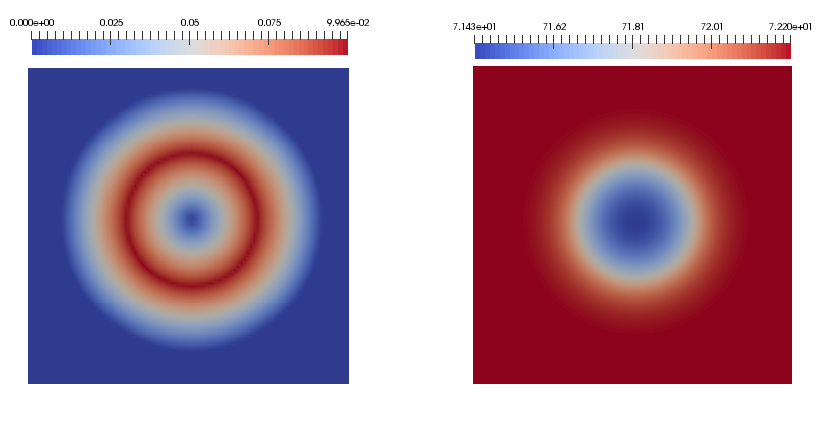}\\
            \caption{Initial Pressure and Mach number distributions for $M_0 = 10^{-1}$.}
\label{fig:1}
  \end{figure}

Initial conditions are given by (see Figure \ref{fig:1} for initial pressure and Mach number distributions for $M_0=10^{-1}$):
\begin{equation}
p_0 = \frac{\rho_0}{\gamma M_0^2}
\end{equation}
\begin{equation}
   (u_{\phi}(r),p(r)) = 
   \begin{cases}  
  		(5r,p_0+\frac{25}{2}r^2),\\ \text{~~~~~~~~~~~~~~~~~~~~~~~if } 0 \leq r < 0.2\\~\\
  		
  		(2-5r, p_0 + \frac{25}{2}r^2 + \\\text{~~~}4(1-5r-\ln 0.2 + \ln 5),\\ \text{~~~~~~~~~~~~~~~~~~~~~if } 0.2 \leq r < 0.4\\~\\
  		
  		(0,p_0 -2 + 4 \ln 2,\\ \text{~~~~~~~~~~~~~~~~~~~~~~~~~~~~if } 0.4 \leq r
\end{cases}
 \end{equation}\\
where $M_0$ is the reference Mach number, and ($r$,$\phi$) are polar coordinates with the origin placed in the center of the vortex.

\begin{figure}
 \centering
		    \includegraphics[width=10.5cm]{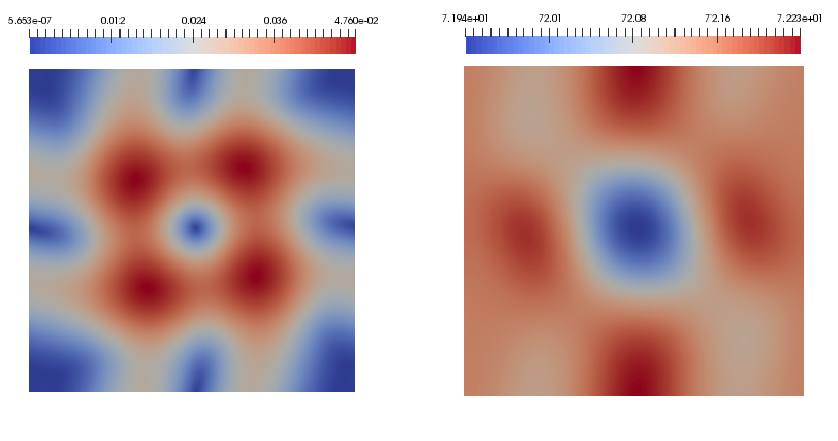}\\
            \caption{Pressure and Mach number distributions at $t=1$ for $M_0 = 10^{-1}$, no splitting error reduction.} 
    \label{fig:2}
  \end{figure}

\begin{figure}[htp]
  \begin{subfigure}[b]{.5\linewidth}
    \centering
    \includegraphics[width=2in]{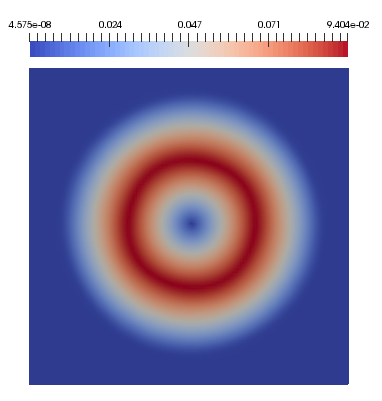}
    \caption{$M_0=10^{-1}$}
    \label{fig:3a}
  \end{subfigure}%
  \begin{subfigure}[b]{.5\linewidth}
    \centering
    \includegraphics[width=2in]{mach1final.png}
    \subcaption{$M_0=10^{-2}$}
    \label{fig:3b}
  \end{subfigure}
   \begin{subfigure}[b]{.5\linewidth}
    \centering
    \includegraphics[width=2in]{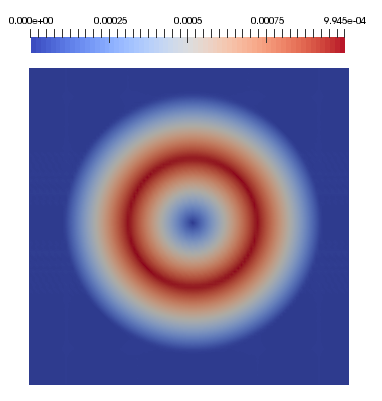}
    \subcaption{$M_0=10^{-3}$}
    \label{fig:3c}
  \end{subfigure}
   \begin{subfigure}[b]{.5\linewidth}
    \centering
    \includegraphics[width=2in]{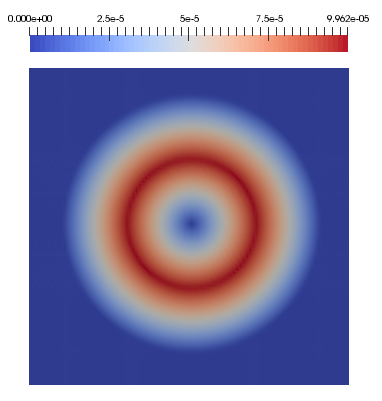}
    \subcaption{$M_0=10^{-4}$}
    \label{fig:3d}
  \end{subfigure}
   \begin{subfigure}[b]{.5\linewidth}
    \centering
    \includegraphics[width=2in]{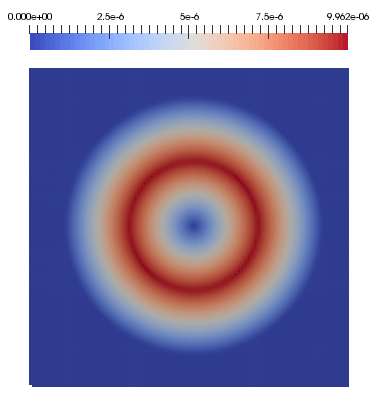}
    \subcaption{$M_0=10^{-5}$}
    \label{fig:3e}
  \end{subfigure}
   \begin{subfigure}[b]{.5\linewidth}
    \centering
    \includegraphics[width=2in]{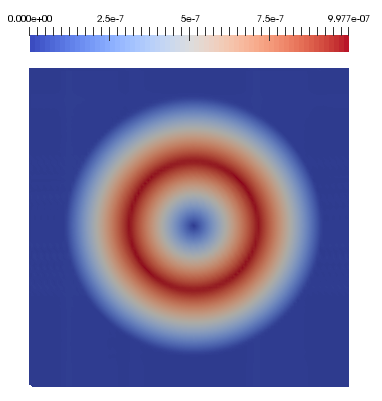}
    \subcaption{$M_0=10^{-6}$}
    \label{fig:3f}
  \end{subfigure}
  \caption{Mach number distributions at t=1 for different $M_0$ computed using the splitting error reduction technique. In all the tests $\tau = 10^{-3}$, grid size: 100x100.}
  \label{fig:3}
\end{figure}

The range of Mach numbers considered here is $[10^{-6};10^{-1}]$. As it can be seen in Figure \ref{fig:2}, if no splitting error reduction technique is used the vortex is not well preserved in time and is completely dissipated after one revolution (by the time $t=1$). However, if one step of the splitting error reduction described in the Section \ref{Splitting error reduction.} is used, the proper time-independent behavior is recovered (see Figure \ref{fig:3}). The method remains efficient in the considered range of Mach numbers, i.e. choosing the reference Mach number as low as $M_0=10^{-6}$ does not require any decrease of time or space steps. Figure \ref{fig:4} shows the time evolution of the relative kinetic energy for the computations described above. It can be concluded that the energy loss is due to the splitting error and its reduction allows one to recover conservative properties up to the influence of the artificial dissipation, required for stability.

All the tests are performed using the LM-dissipation term, described in Section \ref{Damping of high-frequency oscillations.}. Parameters $\omega_1$ and $\omega_2$ are chosen for each $M_0$ to guarantee an acceptable rate of kinetic energy dissipation ($\omega_1$) and dumping of high frequency oscillations ($\omega_2$). Figures \ref{fig:6} and \ref{fig:8} illustrate the effect of these parameters. Figure \ref{fig:7} shows the ineffectiveness of the artificial dissipation operator of the type $\textbf{D}^{LM}_1U$ in dealing with node-to-node pressure oscillations, which have an increasing effect as $M_0$ decreases. Figure \ref{fig:9} reveals that an increase of $\omega_2$ alone leads to an increase of relative kinetic energy and consequently to a rise of instability, especially for relatively high Mach numbers such as $M_0 = 10^{-1}$. Therefore one needs to have both components of the LM-dissipation term. Such combination allows tuning the algorithm for a wide range of Mach numbers. Figure \ref{fig:5} shows that $\frac{|p-p_{max}|}{p_{max}} \sim M^2$, which is the proper scaling in the nearly incompressible regime (see \cite{guil1} for details). Thus, the presented algorithm remains efficient in the low-Mach number regime, since it maintains the stability and accuracy of the simulations without any decrease of spatial or temporal steps, or any other extra computational cost, except explicit error term computations. Numerical oscillations are successfully removed by the proposed stabilization.
  
\pgfplotstableread{greshkin.dat}{\kin}
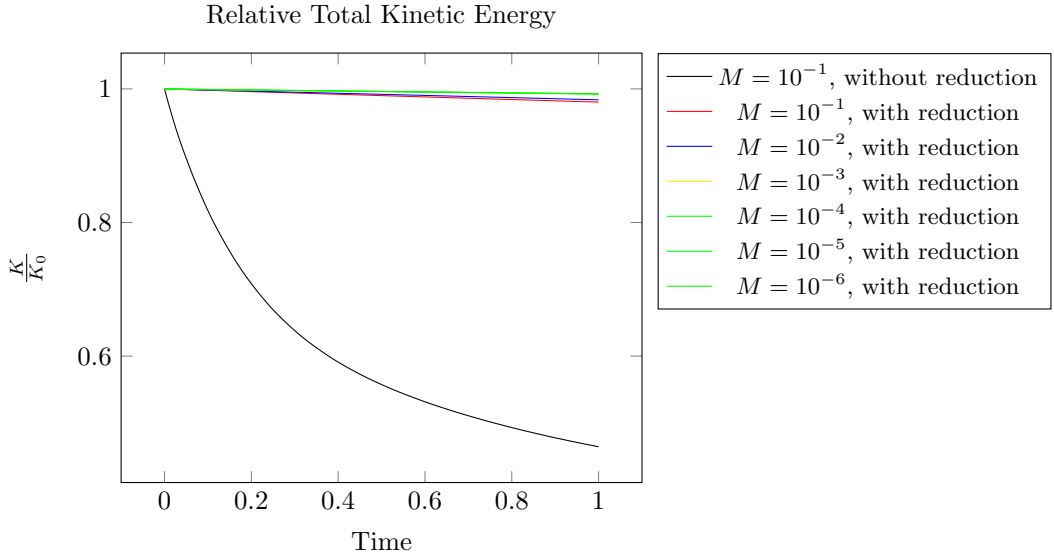
\begin{figure}[htp]
\begin{tikzpicture}[scale=1]
\begin{axis}[legend pos=outer north east,
    title={Relative Total Kinetic Energy},
    xlabel={Time},
    ylabel={$\frac{K}{K_0}$},
]
\addplot [black] table [x={time}, y={k0}] {\kin};
\addlegendentry{\small $M=10^{-1}$, without reduction}
\addplot [red] table [x={time}, y={k1}] {\kin};
\addlegendentry{\small $M=10^{-1}$, with reduction}
\addplot [blue] table [x={time}, y={k2}] {\kin};
\addlegendentry{\small $M=10^{-2}$, with reduction}
\addplot [yellow] table [x={time}, y={k3}] {\kin};
\addlegendentry{\small $M=10^{-3}$, with reduction}
\addplot [green] table [x={time}, y={k4}] {\kin};
\addlegendentry{\small $M=10^{-4}$, with reduction}
\addplot [green] table [x={time}, y={k5}] {\kin};
\addlegendentry{\small $M=10^{-5}$, with reduction}
\addplot [green] table [x={time}, y={k6}] {\kin};
\addlegendentry{\small $M=10^{-6}$, with reduction}
\end{axis}
\path
    ([shift={( 5\pgflinewidth, 5\pgflinewidth)}]current bounding box.north east);
\end{tikzpicture}
\caption{Time evolution of the Relative Kinetic Energy for different test cases. In all the tests $\tau = 10^{-3}$, grid size: 100x100.}
\label{fig:4}
\end{figure}

\pgfplotstableread{pres1.dat}{\presone}
\pgfplotstableread{pres2.dat}{\prestwo}
\pgfplotstableread{pres3.dat}{\presthree}
\pgfplotstableread{pres4.dat}{\presfour}
\pgfplotstableread{pres5.dat}{\presfive}
\pgfplotstableread{pres6.dat}{\pressix}


\begin{figure}[htp]
  \begin{subfigure}[b]{.5\linewidth}
    \centering
\begin{tikzpicture}[scale=0.7]
\begin{axis}[legend pos=outer north east,
    xlabel={x},
    ylabel={$\frac{|p-p_{max}|}{p_{max}}$},	
]
\addplot [black,mark = diamond] table [x={y}, y={presynorm}] {\presone};
\end{axis}
\path
    ([shift={( 5\pgflinewidth, 5\pgflinewidth)}]current bounding box.north east);
\end{tikzpicture}	

    \caption{$M_0=10^{-1}$}
    \label{fig:5a}
  \end{subfigure}%
  \begin{subfigure}[b]{.5\linewidth}
    \centering
    \begin{tikzpicture}[scale=0.7]
\begin{axis}[legend pos=outer north east,
    xlabel={x},
    ylabel={$\frac{|p-p_{max}|}{p_{max}}$},	
]
\addplot [black,mark = diamond] table [x={y}, y={presynorm}] {\prestwo};
\end{axis}
\path
    ([shift={( 5\pgflinewidth, 5\pgflinewidth)}]current bounding box.north east);
\end{tikzpicture}	
    \subcaption{$M_0=10^{-2}$}
    \label{fig:5b}
  \end{subfigure}
   \begin{subfigure}[b]{.5\linewidth}
    \centering
    \begin{tikzpicture}[scale=0.7]
\begin{axis}[legend pos=outer north east,
    xlabel={x},
    ylabel={$\frac{|p-p_{max}|}{p_{max}}$},	
]
\addplot [black,mark = diamond] table [x={y}, y={presynorm}] {\presthree};
\end{axis}
\path
    ([shift={( 5\pgflinewidth, 5\pgflinewidth)}]current bounding box.north east);
\end{tikzpicture}	
    \subcaption{$M_0=10^{-3}$}
    \label{fig:5c}
  \end{subfigure}
   \begin{subfigure}[b]{.5\linewidth}
    \centering
    \begin{tikzpicture}[scale=0.7]
\begin{axis}[legend pos=outer north east,
    xlabel={x},
    ylabel={$\frac{|p-p_{max}|}{p_{max}}$},	
]
\addplot [black,mark = diamond] table [x={y}, y={presynorm}] {\presfour};
\end{axis}
\path
    ([shift={( 5\pgflinewidth, 5\pgflinewidth)}]current bounding box.north east);
\end{tikzpicture}	
    \subcaption{$M_0=10^{-4}$}
    \label{fig:5d}
  \end{subfigure}
   \begin{subfigure}[b]{.5\linewidth}
    \centering
    \begin{tikzpicture}[scale=0.7]
\begin{axis}[legend pos=outer north east,
    xlabel={x},
    ylabel={$\frac{|p-p_{max}|}{p_{max}}$},	
]
\addplot [black,mark = diamond] table [x={y}, y={presynorm}] {\presfive};
\end{axis}
\path
    ([shift={( 5\pgflinewidth, 5\pgflinewidth)}]current bounding box.north east);
\end{tikzpicture}	
    \subcaption{$M_0=10^{-5}$}
    \label{fig:5e}
  \end{subfigure}
   \begin{subfigure}[b]{.5\linewidth}
    \centering
    \begin{tikzpicture}[scale=0.7]
\begin{axis}[legend pos=outer north east,
    xlabel={x},
    ylabel={$\frac{|p-p_{max}|}{p_{max}}$},	
]
\addplot [black,mark = diamond] table [x={y}, y={presynorm}] {\pressix};
\end{axis}
\path
    ([shift={( 5\pgflinewidth, 5\pgflinewidth)}]current bounding box.north east);
\end{tikzpicture}	
    \subcaption{$M_0=10^{-6}$}
    \label{fig:5f}
  \end{subfigure}
  \caption{Relative pressure variations for different $M_0$, cross-sections at $x=0.5$. In all the tests $\tau = 10^{-3}$, grid size: 100x100.}
  \label{fig:5}
\end{figure}

  \pgfplotstableread{kin1.dat}{\kinone}
 \begin{figure}[htp]
\begin{tikzpicture}[scale=1]
\begin{axis}[legend pos=outer north east,
    title={Relative Total Kinetic Energy},
    xlabel={Time},
    ylabel={$\frac{K}{K_0}$},
]
\addplot [black] table [x={time}, y={k1}] {\kinone};
\addlegendentry{$\omega_1=0$}
\addplot [red] table [x={time}, y={k2}] {\kinone};
\addlegendentry{$\omega_1=0.4$}
\addplot [blue] table [x={time}, y={k3}] {\kinone};
\addlegendentry{$\omega_1=0.8$}
\addplot [yellow] table [x={time}, y={k4}] {\kinone};
\addlegendentry{$\omega_1=2$}
\end{axis}
\path
    ([shift={( 5\pgflinewidth, 5\pgflinewidth)}]current bounding box.north east);
\end{tikzpicture}
\caption{Time evolution of the Relative Kinetic Energy depending on the value of $\omega_1$. In all the tests $\omega_2 = 0$, $M_0=10^{-1}$, $\tau = 10^{-3}$, grid size: 100x100.}
\label{fig:6}
\end{figure}

\pgfplotstableread{kin2.dat}{\kintwo}

\begin{figure}[htp]
  \begin{subfigure}[b]{.5\linewidth}
    \centering
\begin{tikzpicture}[scale=0.7]
\begin{axis}[legend pos=outer north east,
    xlabel={x},
    ylabel={$\frac{|p-p_{max}|}{p_{max}}$},	
]
\addplot [black,mark = diamond] table [x={y}, y={presy1norm}] {\kintwo};
\end{axis}
\path
    ([shift={( 5\pgflinewidth, 5\pgflinewidth)}]current bounding box.north east);
\end{tikzpicture}	

    \caption{$\omega_1=0$}
    \label{fig:7a}
  \end{subfigure}%
  \begin{subfigure}[b]{.5\linewidth}
    \centering
    \begin{tikzpicture}[scale=0.7]
\begin{axis}[legend pos=outer north east,
    xlabel={x},
    ylabel={$\frac{|p-p_{max}|}{p_{max}}$},	
]
\addplot [black,mark = diamond] table [x={y}, y={presy2norm}] {\kintwo};
\end{axis}
\path
    ([shift={( 5\pgflinewidth, 5\pgflinewidth)}]current bounding box.north east);
\end{tikzpicture}	
    \subcaption{$\omega_1=0.4$}
    \label{fig:7b}
  \end{subfigure}
   \begin{subfigure}[b]{.5\linewidth}
    \centering
    \begin{tikzpicture}[scale=0.7]
\begin{axis}[legend pos=outer north east,
    xlabel={x},
    ylabel={$\frac{|p-p_{max}|}{p_{max}}$},	
]
\addplot [black,mark = diamond] table [x={y}, y={presy3norm}] {\kintwo};
\end{axis}
\path
    ([shift={( 5\pgflinewidth, 5\pgflinewidth)}]current bounding box.north east);
\end{tikzpicture}	
    \subcaption{$\omega_1=0.8$}
    \label{fig:7c}
  \end{subfigure}
   \begin{subfigure}[b]{.5\linewidth}
    \centering
    \begin{tikzpicture}[scale=0.7]
\begin{axis}[legend pos=outer north east,
    xlabel={x},
    ylabel={$\frac{|p-p_{max}|}{p_{max}}$},	
]
\addplot [black,mark = diamond] table [x={y}, y={presy4norm}] {\kintwo};
\end{axis}
\path
    ([shift={( 5\pgflinewidth, 5\pgflinewidth)}]current bounding box.north east);
\end{tikzpicture}	
    \subcaption{$\omega_1=1$}
    \label{fig:7d}
  \end{subfigure}
   \caption{Relative pressure variation after one time step, cross-section at $x=0.5$, for different values of $\omega_1$. Further integration leads to a numerical failure. In all the tests $\omega_2 = 0$, $M_0=10^{-6}$, $\tau = 10^{-3}$, grid size: 100x100.}
   \label{fig:7}
\end{figure}

\pgfplotstableread{osc1.dat}{\oscone}

\begin{figure}[htp]
  \begin{subfigure}[b]{.5\linewidth}
    \centering
\begin{tikzpicture}[scale=0.7]
\begin{axis}[legend pos=outer north east,
    xlabel={x},
    ylabel={$\frac{|p-p_{max}|}{p_{max}}$},	
]
\addplot [black,mark = diamond] table [x={y}, y={presy1norm}] {\oscone};
\end{axis}
\path
    ([shift={( 5\pgflinewidth, 5\pgflinewidth)}]current bounding box.north east);
\end{tikzpicture}	

    \subcaption{$\omega_2 = 0.01$}
    \label{fig:8a}
  \end{subfigure}%
  \begin{subfigure}[b]{.5\linewidth}
    \centering
    \begin{tikzpicture}[scale=0.7]
\begin{axis}[legend pos=outer north east,
    xlabel={x},
    ylabel={$\frac{|p-p_{max}|}{p_{max}}$},	
]
\addplot [black,mark = diamond] table [x={y}, y={presy2norm}] {\oscone};
\end{axis}
\path
    ([shift={( 5\pgflinewidth, 5\pgflinewidth)}]current bounding box.north east);
\end{tikzpicture}	
    \subcaption{$\omega_2 = 0.1$}
    \label{fig:8b}
  \end{subfigure}
   \begin{subfigure}[b]{.5\linewidth}
    \centering
    \begin{tikzpicture}[scale=0.7]
\begin{axis}[legend pos=outer north east,
    xlabel={x},
    ylabel={$\frac{|p-p_{max}|}{p_{max}}$},	
]
\addplot [black,mark = diamond] table [x={y}, y={presy3norm}] {\oscone};
\end{axis}
\path
    ([shift={( 5\pgflinewidth, 5\pgflinewidth)}]current bounding box.north east);
\end{tikzpicture}	
    \subcaption{$\omega_2=0.5$}
    \label{fig:8c}
  \end{subfigure}
   \begin{subfigure}[b]{.5\linewidth}
    \centering
    \begin{tikzpicture}[scale=0.7]
\begin{axis}[legend pos=outer north east,
    xlabel={x},
    ylabel={$\frac{|p-p_{max}|}{p_{max}}$},	
]
\addplot [black,mark = diamond] table [x={y}, y={presy4norm}] {\oscone};
\end{axis}
\path
    ([shift={( 5\pgflinewidth, 5\pgflinewidth)}]current bounding box.north east);
\end{tikzpicture}	
    \subcaption{$\omega_2 = 1.1$}
    \label{fig:8d}
  \end{subfigure}
 \caption{Relative pressure variations after one time step, cross-section at $x=0.5$, for different values of $\omega_2$. In all the tests $\omega_1 = 1$, $M_0=10^{-6}$, $\tau = 10^{-3}$, grid size: 100x100.}
 \label{fig:8}
\end{figure}

  \pgfplotstableread{osc2.dat}{\osctwo}
  \begin{figure}[htp]
\begin{tikzpicture}[scale=1]
\begin{axis}[legend pos=outer north east,
    title={Relative Total Kinetic Energy},
    xlabel={Time},
    ylabel={$\frac{K}{K_0}$},
]
\addplot [black] table [x={time}, y={k1}] {\osctwo};
\addlegendentry{$\omega_2 = 0$}
\addplot [red] table [x={time}, y={k2}] {\osctwo};
\addlegendentry{$\omega_2=0.1$}
\addplot [blue] table [x={time}, y={k3}] {\osctwo};
\addlegendentry{$\omega_2 = 0.2$}
\addplot [yellow] table [x={time}, y={k4}] {\osctwo};
\addlegendentry{$\omega_2 = 0.22$}
\end{axis}
\path
    ([shift={( 5\pgflinewidth, 5\pgflinewidth)}]current bounding box.north east);
\end{tikzpicture}
\caption{Time evolution of the Relative Kinetic Energy depending on the value of $\omega_2$. In all the tests $\omega_1 = 0$, $M_0=10^{-1}$, $\tau = 10^{-3}$, grid size: 100x100.}
\label{fig:9}
\end{figure}

\subsection{Sod Shock Tube.}
\label{Sod Shock Tube.}

The Sod Shock tube test (originally proposed in \cite{sod}) is used here to evaluate shock capturing properties of the method when the GP-dissipation term is used, and compare the implicit version of the scheme (IGP) with the explicit one (EGP) and with the classic Lax-Friedrichs method (LF).

The initial conditions are given by:
\begin{equation}
\begin{bmatrix}
\rho_L \\ p_L \\ u_L
\end{bmatrix} = 
\begin{bmatrix}
1.0 \\ 1.0 \\ 0.0
\end{bmatrix}
\end{equation}
\begin{equation}
\begin{bmatrix}
\rho_R \\ p_R \\ u_R
\end{bmatrix} = 
\begin{bmatrix}
0.125 \\ 0.1 \\ 0.0
\end{bmatrix}
\end{equation}\\

The results at $T_{end} = 0.1$ for the density ($\rho$), pressure ($p$), velocity ($u$) and internal energy ($e$) are presented in Figure \ref{fig:10} ($h=10^{-3}$) and Figure \ref{fig:11} ($h=0.5 \cdot 10^{-3}$). In both cases $\tau = 10^{-4}$.

It can be seen that explicit and implicit versions of the Guermond-Popov scheme produce almost equivalent solutions in both cases. On a coarser grid, when the space error term has a dominant contribution to the overall error, the Guermond-Popov scheme maintains significantly sharper profiles of discontinuities in the solution than the Lax-Friedrichs method. The results of this test case demonstrate that while the method still adds a first-order dissipation term in its high-Mach number version, the special scaling of the artificial viscosity (proposed in \cite{invdom}) reduces the amount of dissipation introduced by the stabilization term, without loss of robustness.
 
 \pgfplotstableread{spaceerr.dat}{\spaceerr}
  
\begin{figure}[htp]
  \begin{subfigure}[b]{.5\linewidth}
    \centering
\begin{tikzpicture}[scale=0.7]
\begin{axis}[legend pos= north east,
    title={Density},
    xlabel={x},
    ylabel={$\rho$},
]
\addplot [blue] table [x={x}, y={rhoim}] {\spaceerr};
\addlegendentry{IGP}
\addplot [green] table [x={x}, y={rhoexp}] {\spaceerr};
\addlegendentry{EGP}
\addplot [red] table [x={x}, y={rholf}] {\spaceerr};
\addlegendentry{LF}
\addplot [black,dashed] table [x={x}, y={rhoexact}] {\spaceerr};
\addlegendentry{exact}
\end{axis}
\path
    ([shift={( 5\pgflinewidth, 5\pgflinewidth)}]current bounding box.north east);
\end{tikzpicture}
    \label{fig:10a}
  \end{subfigure}%
  \begin{subfigure}[b]{.5\linewidth}
    \centering
  \begin{tikzpicture}[scale=0.7]
\begin{axis}[legend pos=north east,
    title={Pressure},
    xlabel={x},
    ylabel={$p$},
]
\addplot [blue] table [x={x}, y={pim}] {\spaceerr};
\addlegendentry{IGP}
\addplot [green] table [x={x}, y={pexp}] {\spaceerr};
\addlegendentry{EGP}
\addplot [red] table [x={x}, y={plf}] {\spaceerr};
\addlegendentry{LF}
\addplot [black,dashed] table [x={x}, y={pexact}] {\spaceerr};
\addlegendentry{exact}
\end{axis}
\path
    ([shift={( 5\pgflinewidth, 5\pgflinewidth)}]current bounding box.north east);
\end{tikzpicture}
    \label{fig:10b}
  \end{subfigure}
   \begin{subfigure}[b]{.5\linewidth}
    \centering
    \begin{tikzpicture}[scale=0.7]
\begin{axis}[legend pos=north west,
    title={Velocity},
    xlabel={x},
    ylabel={$u$},
]
\addplot [blue] table [x={x}, y={uim}] {\spaceerr};
\addlegendentry{IGP}
\addplot [green] table [x={x}, y={uexp}] {\spaceerr};
\addlegendentry{EGP}
\addplot [red] table [x={x}, y={ulf}] {\spaceerr};
\addlegendentry{LF}
\addplot [black,dashed] table [x={x}, y={uexact}] {\spaceerr};
\addlegendentry{exact}
\end{axis}
\path
    ([shift={( 5\pgflinewidth, 5\pgflinewidth)}]current bounding box.north east);
\end{tikzpicture}
    \label{fig:10c}
  \end{subfigure}
     \begin{subfigure}[b]{.5\linewidth}
    \centering
    \begin{tikzpicture}[scale=0.7]
\begin{axis}[legend pos=north east,
    title={Internal Energy},
    xlabel={x},
    ylabel={$e$},
]
\addplot [blue] table [x={x}, y={eim}] {\spaceerr};
\addlegendentry{IGP}
\addplot [green] table [x={x}, y={eexp}] {\spaceerr};
\addlegendentry{EGP}
\addplot [red] table [x={x}, y={elf}] {\spaceerr};
\addlegendentry{LF}
\addplot [black,dashed] table [x={x}, y={eexact}] {\spaceerr};
\addlegendentry{exact}
\end{axis}
\path
    ([shift={( 5\pgflinewidth, 5\pgflinewidth)}]current bounding box.north east);
\end{tikzpicture}
    \label{fig:10d}
      \end{subfigure}
  \caption{Sod Shock Tube test, $T_{end} = 10^{-1}$, $h = 10^{-3}$}
  \label{fig:10}
\end{figure}

  \pgfplotstableread{timeerr.dat}{\timeerr}
\begin{figure}[htp]
  \begin{subfigure}[b]{.5\linewidth}
    \centering
\begin{tikzpicture}[scale=0.7]
\begin{axis}[legend pos= north east,
    title={Density},
    xlabel={x},
    ylabel={$\rho$},
]
\addplot [blue] table [x={x}, y={rhoim}] {\timeerr};
\addlegendentry{IGP}
\addplot [green] table [x={x}, y={rhoexp}] {\timeerr};
\addlegendentry{EGP}
\addplot [red] table [x={x}, y={rholf}] {\timeerr};
\addlegendentry{LF}
\addplot [black,dashed] table [x={x}, y={rhoexact}] {\timeerr};
\addlegendentry{exact}
\end{axis}
\path
    ([shift={( 5\pgflinewidth, 5\pgflinewidth)}]current bounding box.north east);
\end{tikzpicture}
    \label{fig:11a}
  \end{subfigure}%
  \begin{subfigure}[b]{.5\linewidth}
    \centering
  \begin{tikzpicture}[scale=0.7]
\begin{axis}[legend pos=north east,
    title={Pressure},
    xlabel={x},
    ylabel={$p$},
]
\addplot [blue] table [x={x}, y={pim}] {\timeerr};
\addlegendentry{IGP}
\addplot [green] table [x={x}, y={pexp}] {\timeerr};
\addlegendentry{EGP}
\addplot [red] table [x={x}, y={plf}] {\timeerr};
\addlegendentry{LF}
\addplot [black,dashed] table [x={x}, y={pexact}] {\timeerr};
\addlegendentry{exact}
\end{axis}
\path
    ([shift={( 5\pgflinewidth, 5\pgflinewidth)}]current bounding box.north east);
\end{tikzpicture}
    \label{fig:11b}
  \end{subfigure}
   \begin{subfigure}[b]{.5\linewidth}
    \centering
    \begin{tikzpicture}[scale=0.7]
\begin{axis}[legend pos=north west,
    title={Velocity},
    xlabel={x},
    ylabel={$u$},
]
\addplot [blue] table [x={x}, y={uim}] {\timeerr};
\addlegendentry{IGP}
\addplot [green] table [x={x}, y={uexp}] {\timeerr};
\addlegendentry{EGP}
\addplot [red] table [x={x}, y={ulf}] {\timeerr};
\addlegendentry{LF}
\addplot [black,dashed] table [x={x}, y={uexact}] {\timeerr};
\addlegendentry{exact}
\end{axis}
\path
    ([shift={( 5\pgflinewidth, 5\pgflinewidth)}]current bounding box.north east);
\end{tikzpicture}
    \label{fig:11c}
  \end{subfigure}
     \begin{subfigure}[b]{.5\linewidth}
    \centering
    \begin{tikzpicture}[scale=0.7]
\begin{axis}[legend pos=north east,
    title={Internal Energy},
    xlabel={x},
    ylabel={$e$},
]
\addplot [blue] table [x={x}, y={eim}] {\timeerr};
\addlegendentry{IGP}
\addplot [green] table [x={x}, y={eexp}] {\timeerr};
\addlegendentry{EGP}
\addplot [red] table [x={x}, y={elf}] {\timeerr};
\addlegendentry{LF}
\addplot [black,dashed] table [x={x}, y={eexact}] {\timeerr};
\addlegendentry{exact}
\end{axis}
\path
    ([shift={( 5\pgflinewidth, 5\pgflinewidth)}]current bounding box.north east);
\end{tikzpicture}
    \label{fig:11d}
  \end{subfigure}
   \caption{Sod Shock Tube test, $T_{end} = 10^{-1}$, $h = 0.5 \cdot 10^{-3}$}
   \label{fig:11}
\end{figure}
\subsection{Manufactured Solution.}
\label{Manufactured Solution.}
The method of Manufactured Solution (see \cite{manuf}) is used here to verify the numerical code implementation and convergence properties of the proposed algorithm. The idea of the method is to choose an analytical solution and modify the governing equations by the inclusion of source terms, computed using the solution. Here, we define primitive variables ($\rho$,$u$,$v$, and $p$) as:\\
\begin{equation}
\rho(x,y,t) = \rho_0 + \rho_t \sin{t} + \rho_x \sin{x} + \rho_y \cos{y}
\end{equation}
\begin{equation}
u(x,y,t) = u_0 + u_t \sin{t} + u_x \sin{x} + u_y \cos{y}
\end{equation}
\begin{equation}
v(x,y,t) = v_0 + v_t \cos{t} + v_x \cos{x} + v_y \sin{y}
\end{equation}
\begin{equation}
p(x,y,t) = p_0 + p_t \cos{t} + p_x \cos{x} + p_y \sin{y},
\end{equation}\\
where $\xi_0$, $\xi_t$, $\xi_x$, $\xi_y$ (for $\xi$ = $\rho$, $u$, $v$, or $p$) are constants defined separately for high- ($M_0 = 2.14$) and low- ($M_0 = 2.14 \cdot 10^{-4}$) Mach number cases in Tables \ref{tab:2} and \ref{tab:3}. Conservative variables are computed from the primitive ones as usual. In both cases the single fluid ideal gas equation of state is used ($\gamma=1.4$, $\pi^{\infty} = 0$) with viscosity parameter $\mu = 1$ $[kg/(s \cdot m)]$. The low-Mach number stabilization term $\textbf{D}^{LM}U$ is scaled with the parameters $\omega_1 = 1.1$, and $\omega_2 = 10^{-2}$. A single-step version of the splitting error reduction procedure was used for both, high- and low- Mach number cases.
\newpage
High-Mach number computations are performed on a [$1 \times 1$] square domain with exact Dirichlet boundary conditions:
\begin{equation}
\hat{U}^{n+1}= (\textbf{I}+\textbf{A}_y)U_{exact}^{n+1}
\end{equation}
\begin{equation}
U^{n+1} = U_{exact}^{n+1}. 
\end{equation}
In the low-Mach number case, a square [$2\pi \times 2\pi$] domain with periodic boundary conditions is used.

Figures \ref{fig:13} and \ref{fig:14} show the expected order of accuracy for smooth solutions (first-order convergence in time for both versions of the scheme, first-order convergence in space for the high-Mach number version, and second-order convergence in space for the low-Mach number version). Since the algorithm remains accurate, stable, and  oscillation-free for the low-Mach number case, it is demonstrated to be applicable for time-dependent viscous nearly incompressible flows.
\begin{table}
\begin{tabular}{ |p{2cm}||p{2cm}|p{2cm}||p{2cm}||p{2cm}|  }
 \hline
 Equation, $\xi$ & $\xi_0$ & $\xi_t$ & $\xi_x$ & $\xi_y$\\
 \hline
  $\rho$ $\left ( kg/m^3 \right )$& $1$ &$0.5$&$0.015$&$-0.01$\\
 $u$ $\left (m/s \right )$&   $8$& $4$& $0.05$& $-0.03$ \\
 $v$ $\left (m/s \right )$&$8$ & $4$ & $-0.075$ & $0.04$\\
 $p$ $\left (N/m^2 \right )$&$10$ & $1$ & $0.02$ & $0.05$\\
 \hline
\end{tabular}
\caption{Constants for high-Mach number Manufactured Solution.}
\label{tab:2}
\end{table}

\begin{table}
\begin{tabular}{ |p{2cm}||p{2cm}|p{2cm}||p{2cm}||p{2cm}|  }
 \hline
 Equation, $\xi$ & $\xi_0$ & $\xi_t$ & $\xi_x$ & $\xi_y$\\
 \hline
  $\rho$ $\left ( kg/m^3 \right )$& $1$ &$0.5$&$0.015$&$-0.01$\\ 
 $u$ $\left (m/s \right )$&  $8 \cdot 10^{-4}$& $4 \cdot 10^{-4}$& $0.05 \cdot 10^{-4}$& $-0.03 \cdot 10^{-4}$ \\
 $v$ $\left (m/s \right )$&$8 \cdot 10^{-4}$ & $4 \cdot 10^{-4}$ & $-0.075 \cdot 10^{-4}$ & $0.04 \cdot 10^{-4}$\\
 $p$ $\left (N/m^2 \right )$&$10$ & $1$ & $0.02$ & $0.05$\\
 \hline
\end{tabular}
\caption{Constants for low-Mach number Manufactured Solution.}
\label{tab:3}
\end{table}
\pgfplotstableread{spaceHM.dat}{\shm}
\pgfplotstableread{timeHM.dat}{\thm}
\pgfplotstableread{spaceLM.dat}{\slm}
\pgfplotstableread{timeLM.dat}{\tlm}\
\begin{figure}[htp] 
  \begin{subfigure}[b]{.5\linewidth}
    \centering
    \begin{tikzpicture}[scale=0.7]
\begin{axis}[legend pos=north west,
xlabel={$\ln{h}$},
    ylabel={$\ln{(L^1\text{error})}$}
]
\addplot [black,mark = diamond] table [x={lnh}, y={lnEl1}] {\shm};
\addlegendentry{$\ln{(L^1\text{error})}$}
\addplot [blue,dashed] table [x={lnh}, y={lnhmod}] {\shm};
\addlegendentry{1st-order slope}
\end{axis}
\path
    ([shift={( 5\pgflinewidth, 5\pgflinewidth)}]current bounding box.north east);
\end{tikzpicture}	
    \label{fig:13a}
  \end{subfigure}
     \begin{subfigure}[b]{.5\linewidth}
    \centering
\begin{tikzpicture}[scale=0.7]
\begin{axis}[legend pos=north west,
xlabel={$\ln{h}$},
     ylabel={$\ln{(L^1\text{error})}$}
]
\addplot [black,mark = diamond] table [x={lnh}, y={lnEl1}] {\slm};
\addlegendentry{$\ln{(L^1\text{error})}$}
\addplot [blue,dashed] table [x={lnh}, y={lnhmod}] {\slm};
\addlegendentry{1st-order slope}
\addplot [red,dashed] table [x={lnh}, y={2lnh}] {\slm};
\addlegendentry{2st-order slope}
\end{axis}
\path
    ([shift={( 5\pgflinewidth, 5\pgflinewidth)}]current bounding box.north east);
\end{tikzpicture}	
   \label{fig:13b}
\end{subfigure}
  \caption{$log$-$log$ plots of the discrete $L^{1}$ norm of the total energy errors at $t=10^{-2}$ ($\tau = 10^{-4}$) for high-Mach number (left) and low-Mach number (right) manufactured solutions.}
  \label{fig:13}
  \end{figure}
  \begin{figure}[htp]
 \begin{subfigure}[b]{.5\linewidth}
    \centering
    \begin{tikzpicture}[scale=0.7]
\begin{axis}[legend pos=north west,
	xlabel={$\ln{\tau}$},
     ylabel={$\ln{(L^1\text{error})}$}
]
\addplot [black,mark = diamond] table [x={lntau}, y={lnEl1}] {\thm};
\addlegendentry{$\ln{(L^1\text{error})}$}
\addplot [blue,dashed] table [x={lntau}, y={lntau}] {\thm};
\addlegendentry{1st-order slope}
\end{axis}
\path
    ([shift={( 5\pgflinewidth, 5\pgflinewidth)}]current bounding box.north east);
\end{tikzpicture}	
    \label{fig:14a}
  \end{subfigure}
  \begin{subfigure}[b]{.5\linewidth}
    \centering
    \begin{tikzpicture}[scale=0.7]
\begin{axis}[legend pos=north west,
xlabel={$\ln{\tau}$},
     ylabel={$\ln{(L^1\text{error})}$}
]
\addplot [black,mark = diamond] table [x={lntau}, y={lnEl1}] {\tlm};
\addlegendentry{$\ln{(L^1\text{error})}$}
\addplot [blue,dashed] table [x={lntau}, y={lntaumod}] {\tlm};
\addlegendentry{1st-order slope}
\end{axis}
\path
    ([shift={( 5\pgflinewidth, 5\pgflinewidth)}]current bounding box.north east);
\end{tikzpicture}	
    \label{fig:14b}
  \end{subfigure}
 \caption{$log$-$log$ plots of the discrete $L^{1}$ norm of the total energy errors  at $t=1$ on $322 \times 322$ uniform grid for high-Mach number (left) and low-Mach number (right) manufactured solutions.}
  \label{fig:14}
\end{figure}
\subsection{Weak Scalability.}
\label{Weak Scalability.}
A weak scalability test is provided here to demonstrate the performance of the algorithm on parallel machines. Similar to \cite{minevadi}, a fixed number of grid points per CPU core is considered ($10^6$). Then, CPU-time is recorded for an increasing number of cores. These times, computed while solving the Manufactured Solution test in the high-Mach number regime, can be found in Table \ref{tab:4}. Scaling efficiency, computed as the ratio of the CPU-time on one core to the CPU-time on $n$ cores, is also presented in Table \ref{tab:4}. Taking into account that no attempts have been made to optimize the code in general, or the interprocessor communications in particular, the algorithm shows reasonable weak scalability and thus can be considered as a promising one for parallel computations. All these computations were performed on the GRAHAM cluster provided by Compute Canada (www.computecanada.ca).
\begin{table}
\begin{tabular}{ |p{2cm}||p{2cm}|p{2cm}|p{2cm}|p{2cm}|p{2cm}|  }
 \hline
 \# cores & 1(1$\times$1) & 64(8$\times$8) & 256(16$\times$16) & 512(16$\times$32)&992(32$\times$31)\\
 \hline
  Time&  31.25 $s$ &36.78 $s$&39.12 $s$& 41.85 $s$&45.77 $s$\\ 
  Efficiency&-& 85 \% & 80 \%&  75\%& 68\% \\
 \hline
\end{tabular}
\caption{Weak Scalability test, $10^6$ grin points per core.}
\label{tab:4}
\end{table} 
\subsection{One-Dimensional Interface Advection.}
\label{Interface Advection. 1D.}
Here the advection of an interface between air ($\gamma=1.4$, $\pi^{\infty}=0$) and water ($\gamma=6.12$, non-dimensiolized $\pi^{\infty} = 0.1631$) under the atmospheric pressure is considered (similar to \cite{wenocol}). The purpose of the test is to demonstrate the absence of spurious pressure oscillations and preservation of pressure and velocity equilibrium for the method with both GP- and LM- ($\omega_1=10$, $\omega_2=1$) dissipation terms.

The initial conditions are given by:\\
\begin{equation}
   (\rho,u,p,\phi) = 
   \begin{cases}  
  		(1,0.1,4.819 \times 10^{-5},1), \text{~~~~~~~~~~~~~~~~~~~~~~if }  x 
  		\leq 0.3\\
  		(1.204 \times 10^{-3},0.1,4.819 \times 10^{-5},0), \text{~~~~~~~if } x > 0.3,
\end{cases}
 \end{equation}\\
and are regularized after that by performing the interface sharpening procedure until convergence.
It is evolved with $\tau = 10^{-2}$ on a uniform grid of 130 cells until $t_{end} = 4$. The results and initial conditions are shown in Figure \ref{fig:15}.
\pgfplotstableread{adv.dat}{\adv}

\begin{figure}[htp]
  \begin{subfigure}[b]{.5\linewidth}
    \centering
\begin{tikzpicture}[scale=0.7]
\begin{axis}[legend style={at={(0.5,1.1)},anchor=south},
    xlabel={x},
    ylabel={$\frac{|p-p_{0}|}{p_{0}}$}
]
\addplot [black,mark = diamond] table [x={x}, y={pinnorm}] {\adv};
\addlegendentry{\small $t=0$}
\addplot [black,mark = diamond] table [x={x}, y={phmnorm}] {\adv};
\addlegendentry{\small $t=4$, GP-dissipation term}
\addplot [green,mark = o] table [x={x}, y={plmnorm}] {\adv};
\addlegendentry{\small $t=4$, LM-dissipation term}
\end{axis}
\path
    ([shift={( 5\pgflinewidth, 5\pgflinewidth)}]current bounding box.north east);
\end{tikzpicture}	

    \subcaption{Relative Pressure Error}
    \label{fig:15a}
  \end{subfigure}%
  \begin{subfigure}[b]{.5\linewidth}
    \centering
    \begin{tikzpicture}[scale=0.7]
\begin{axis}[legend style={at={(0.5,1.1)},anchor=south},
    xlabel={x},
    ylabel={$\frac{|u-u_{0}|}{u_{0}}$}
]
\addplot [black,mark = diamond] table [x={x}, y={uinnorm}] {\adv};
\addlegendentry{\small $t=0$}
\addplot [black,mark = diamond] table [x={x}, y={uhmnorm}] {\adv};
\addlegendentry{\small $t=4$, GP-dissipation}
\addplot [green,mark = o] table [x={x}, y={ulmnorm}] {\adv};
\addlegendentry{\small $t=4$, LM-dissipation}
\end{axis}
\path
    ([shift={( 5\pgflinewidth, 5\pgflinewidth)}]current bounding box.north east);
\end{tikzpicture}	
    \subcaption{Relative Velocity Error}
    \label{fig:15b}
  \end{subfigure}
   \begin{subfigure}[b]{.5\linewidth}
    \centering
    \begin{tikzpicture}[scale=0.7]
\begin{axis}[legend style={at={(0.5,-0.2)},anchor=north},
    xlabel={x},
    ylabel={$\rho$}
]
\addplot [black,mark = diamond] table [x={x}, y={rhoin}] {\adv};
\addlegendentry{\small $t=0$}
\addplot [black,mark = diamond] table [x={x}, y={rhohm}] {\adv};
\addlegendentry{\small $t=4$, GP-dissipation}
\addplot [green,mark = o] table [x={x}, y={rholm}] {\adv};
\addlegendentry{\small $t=4$, LM-dissipation}
\end{axis}
\path
    ([shift={( 5\pgflinewidth, 5\pgflinewidth)}]current bounding box.north east);
\end{tikzpicture}	
    \subcaption{Density}
    \label{fig:15c}
  \end{subfigure}
   \begin{subfigure}[b]{.5\linewidth}
    \centering
    \begin{tikzpicture}[scale=0.7]
\begin{axis}[legend style={at={(0.5,-0.2)},anchor=north},
    xlabel={x},
    ylabel={$\phi$}
]
\addplot [black,mark = diamond] table [x={x}, y={phin}] {\adv};
\addlegendentry{\small $t=0$};
\addplot [black,mark = diamond] table [x={x}, y={phihm}] {\adv};
\addlegendentry{\small $t=4$, GP-dissipation};
\addplot [green,mark = o] table [x={x}, y={philm}] {\adv};
\addlegendentry{\small $t=4$, LM-dissipation}
\end{axis}
\path
    ([shift={( 5\pgflinewidth, 5\pgflinewidth)}]current bounding box.north east);
\end{tikzpicture}	
    \subcaption{VoF-function}
    \label{fig:15d}
  \end{subfigure}
\caption{Interface advection using GP- and LM- artificial dissipation terms}
\label{fig:15}
\end{figure}
In both cases (LM- and GP-dissipation terms) contact discontinuity is well preserved and no spurious oscillations are introduced, despite the high density ratio. The sharp profile of the interface is well maintained. Hence, the test confirms the analysis of the contact discontinuities preservation property of the method, and the compatibility of the presented formulation of the VoF method and interface sharpening with the discretization strategy and proposed stabilization terms.

\subsection{Two-Dimensional Interface Advection.}
\label{Interface Advection. 2D.}
The purpose of this test case is to verify the performance of the interface capturing algorithm in two dimensions. A water drop of radius $0.15$ is considered. It is located in the lower left corner of a square $[1 \times 1]$ domain at $(0.3,0.3)$ at the initial moment of time and surrounded by air (both water and air have the same parameters as in the previous test). The bubble is then advected to the upper right corner of the domain, $u = v = 0.1$. The problem is discretized with $\tau = 10^{-2}$ on a $200 \times 200$ uniform grid. No splitting error reduction was needed for this test. LM-dissipation was used ($\omega_1 = 10$, $\omega_2=0$).
 \begin{figure}
 \centering
		    \includegraphics[width=10.5cm]{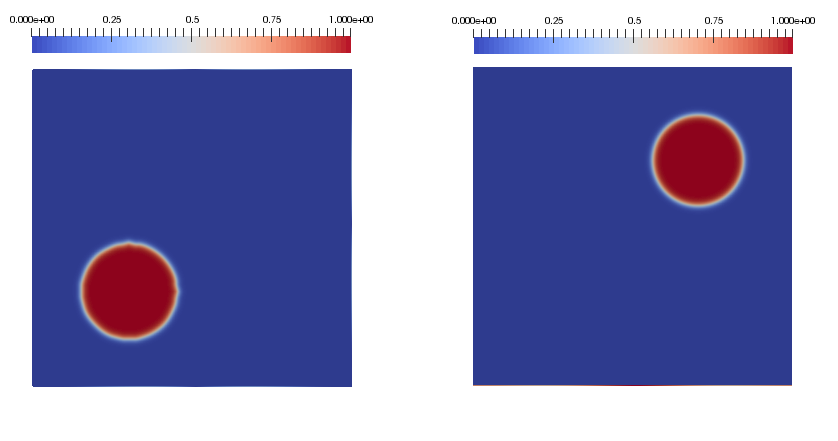}\\
            \caption{\small VoF-function at $t=0$ (left) and $t=4$ (right)}
\label{fig:16}
  \end{figure}

\pgfplotstableread{cross.dat}{\cross}

\begin{figure}[htp]
  \begin{subfigure}[b]{.5\linewidth}
    \centering
\begin{tikzpicture}[scale=0.7]
\begin{axis}[legend style={at={(0.5,1.1)},anchor=south},
    xlabel={x},
    ylabel={$\frac{|p-p_{max}|}{p_{max}}$}
]
\addplot [black,mark = diamond] table [x={x}, y={pinnorm}] {\cross};
\addlegendentry{$t=0$, $y=0.3$}
\addplot [green,mark = o] table [x={x}, y={pendnorm}] {\cross};
\addlegendentry{$t=4$, $y=0.7$}
\end{axis}
\path
    ([shift={( 5\pgflinewidth, 5\pgflinewidth)}]current bounding box.north east);
\end{tikzpicture}	

    \subcaption{Pressure}
    \label{fig:17a}
  \end{subfigure}%
  \begin{subfigure}[b]{.5\linewidth}
    \centering
    \begin{tikzpicture}[scale=0.7]
\begin{axis}[legend style={at={(0.5,1.1)},anchor=south},
    xlabel={x},
    ylabel={$\frac{|u-u_{0}|}{u_{0}}$},	
]
\addplot [black,mark = diamond] table [x={x}, y={uinnorm}] {\cross};
\addlegendentry{$t=0$, $y=0.3$}
\addplot [green,mark = o] table [x={x}, y={uendnorm}] {\cross};
\addlegendentry{$t=4$, $y = 0.7$}
\end{axis}
\path
    ([shift={( 5\pgflinewidth, 5\pgflinewidth)}]current bounding box.north east);
\end{tikzpicture}	
    \subcaption{Velocity}
    \label{fig:17b}
  \end{subfigure}
  \caption{Relative Pressure and Velocity Errors, cross-sections at $y=0.3$ and  $y = 0.7$}
 \label{fig:17}
\end{figure}
As can be seen in Figures \ref{fig:16} and \ref{fig:17}, the shape of the bubble is well preserved and no spurious oscillations or artificial acoustic waves are introduced. Thus, the preservation of contact discontinuities property of the method extends to the multidimensional case as expected. Furthermore, the test shows the ability of the interface-capturing methodology chosen for the scheme to handle nontrivial interface configurations when used in combination with the proposed discretization technique.

\subsection{Laplace formula.}
\label{Laplace formula.}
Following \cite{sten} and \cite{st2}, the implementation of surface tension is verified by reproducing the pressure jump across a curved interface, given by the Laplace formula. In case of a cylindrical interface:
\begin{equation}
\Delta  p_{exact}= \sigma/R,
\end{equation}
where $R$ is the radius of the cylinder.

At the initial moment of time, a liquid ($\rho_l$, $\gamma=2.4$, $\pi^{\infty} = 10^7$  $Pa$) cylindrical drop of radius $R=0.3$ $m$ centered at $(0.5$ $m,0.5$ $m)$ is placed in a $[1m \times 1m]$ square domain with symmetry boundary conditions, surrounded by gas ($\rho_g = 1$ $kg/m^3$, $\gamma=1.4$, $\pi^{\infty} = 0$ $Pa$). The pressure is given by the Laplace formula, with the pressure outside of the drop being $p_{out} = 100$ $Pa$, and the pressure inside being $p_{in} = p_{out} + \Delta p_{exact}$.

The relative pressure jump error $E(\Delta p) = \frac{\Delta p - \Delta p_{exact}}{\Delta p_{exact}}$, and the radius of the cylindrical drop $R$ are computed by an averaging procedure, similar to the one used in \cite{sten}. Nodes with $\phi \geq 0.9$ are considered to be inside, and nodes with $\phi < 0.9$ to be outside of the drop. As in \cite{st2}, spatial convergence of the method is evaluated by measuring $E(\Delta p)$ after one time step for different grid sizes. No splitting error reduction was performed for this test. Results are shown in Figure \ref{fig:18a}.

Figure \ref{fig:18b} shows the increase in discrete maximum norm of the velocity (i.e. an increase of magnitude of parasitic current) under grid refinement. A similar effect was described in \cite{st2} for some other implementations of the CSF approach. No quantitative results on parasitic currents for the method were presented in the original paper \cite{sten} to compare with, thus their influence on accuracy and stability of the method remains an open question. Figures \ref{fig:19}-\ref{fig:20} demonstrate the dependence of pressure and velocity errors after one time step on the values of $\sigma$ and $\rho_l$. The long-time evolution of $E(\Delta p)$ for $\rho_l=10$ $kg/m^3$, $\sigma=1$ $N/m$, and different values of $h_x=h_y=h$ is shown in Figure \ref{fig:21}. The thickness of the interface is also expected to play a role in the accuracy of the approximation. As is shown in Table \ref{tab:1}, a thinner interface (smaller value of $\epsilon$) leads to a smaller error in the pressure jump.

For all values of the parameters considered above, errors remain small. The method is shown to be convergent in space in terms of the pressure error, thus the Laplace law is reproduced by the algorithm. Further investigation of parasitic currents and the influence of various parameters of the problem, as well as the amount of artificial dissipation, on long-time accuracy and stability of the method is an interesting problem. Therefore, the test demonstrated a potential of coupling the surface tension formulation from \cite{sten} with the interface sharpening technique from \cite{sharp}, as well as a need for more detailed study of its properties and the effects of its coupling with various discretization and stabilization methods, and interface treatments under different conditions.\\ \\

\pgfplotstableread{st1.dat}{\sto}
\begin{figure}
\begin{subfigure}[b]{.5\linewidth}
    \centering
\begin{tikzpicture}[scale=0.7]
\begin{axis}[legend pos=south east,
	title={Relative Pressure Jump Error},
    xlabel={$h$, $m$},
    ylabel={$E(\Delta p)$},
]
\addplot [black,mark=diamond] table [x={h}, y={ptotal}] {\sto};
\end{axis}
\path
    ([shift={( 5\pgflinewidth, 5\pgflinewidth)}]current bounding box.north east);
\end{tikzpicture}
\subcaption{\label{fig:18a}}
\end{subfigure}
\begin{subfigure}[b]{.5\linewidth}
    \centering
\begin{tikzpicture}[scale=0.7]
\begin{axis}[legend pos=north east,
    title={Maximum Velocity Error},
    xlabel={$h$, $m$},
    ylabel={$||u||_{\infty}$},
]
\addplot [black,mark=diamond] table [x={h}, y={umax}] {\sto};
\end{axis}
\path
    ([shift={( 5\pgflinewidth, 5\pgflinewidth)}]current bounding box.north east);
\end{tikzpicture}
\subcaption{\label{fig:18b}}
\end{subfigure}
\caption{\small $E(\Delta p)$ and $||u||_{\infty}$ after one time step on different grids. $\sigma = 1$
$N/m$,  $\rho_l = 100$ $kg/m^3$, $\tau = 10^{-5}$ $s$.}
\label{fig:18}
\end{figure}

\pgfplotstableread{st2.dat}{\stt}
\begin{figure}
\begin{subfigure}[b]{.5\linewidth}
    \centering
\begin{tikzpicture}[scale=0.7]
\begin{axis}[legend pos=north east,
    title={Relative Pressure Jump Error},
    xlabel={$\sigma$, $N/m$},
    ylabel={$E(\Delta p)$},
]
\addplot [black,mark=diamond] table [x={sigma}, y={ptotal}] {\stt};
\end{axis}
\path
    ([shift={( 5\pgflinewidth, 5\pgflinewidth)}]current bounding box.north east);
\end{tikzpicture}
\subcaption{\label{fig:19a}}
\end{subfigure}
\begin{subfigure}[b]{.5\linewidth}
    \centering
\begin{tikzpicture}[scale=0.7]
\begin{axis}[legend pos=south east,
    title={Maximum Velocity Error},
    xlabel={$\sigma$, $N/m$},
    ylabel={$||u||_{\infty}$},
]
\addplot [black,mark=diamond] table [x={sigma}, y={umax}] {\stt};
\end{axis}
\path
    ([shift={( 5\pgflinewidth, 5\pgflinewidth)}]current bounding box.north east);
\end{tikzpicture}
\subcaption{\label{fig:19b}}
\end{subfigure}
\caption{\small $E(\Delta p)$ and $||u||_{\infty}$ after one time step for different $\sigma$. $h = 5.6 \cdot 10^{-3}$ $m$, $\rho_l = 100$ $kg/m^3$, $\tau = 10^{-5}$ $s$.}
\label{fig:19}
\end{figure}

\pgfplotstableread{st3.dat}{\stth}
\begin{figure}
\begin{subfigure}[b]{.5\linewidth}
    \centering
\begin{tikzpicture}[scale=0.7]
\begin{axis}[legend pos=north east,
    title={Relative Pressure Jump Error},
    xlabel={$\rho_l$, $kg/m^3$},
    ylabel={$E(\Delta p)$},
]
\addplot [black,mark=diamond] table [x={rho}, y={ptotal}] {\stth};
\end{axis}
\path
    ([shift={( 5\pgflinewidth, 5\pgflinewidth)}]current bounding box.north east);
\end{tikzpicture}
\subcaption{\label{fig:20a}}
\end{subfigure}
\begin{subfigure}[b]{.5\linewidth}
    \centering
\begin{tikzpicture}[scale=0.7]
\begin{axis}[legend pos=north east,
    title={Maximum Velocity Error},
    xlabel={$\rho_l$, $kg/m^3$},
    ylabel={$||u||_{\infty}$},
]
\addplot [black,mark=diamond] table [x={rho}, y={umax}] {\stth};
\end{axis}
\path
    ([shift={( 5\pgflinewidth, 5\pgflinewidth)}]current bounding box.north east);
\end{tikzpicture}
\subcaption{\label{fig:20b}}
\end{subfigure}
\caption{\small $E(\Delta p)$ and $||u||_{\infty}$ after one time step for different $\rho_l$. $h = 5.6 \cdot 10^{-3}$ $m$, $\sigma = 1$ $N/m$, $\tau = 10^{-5}$ $s$.}
\label{fig:20}
\end{figure}

\pgfplotstableread{sta.dat}{\sta}
\begin{figure}[htp]
\begin{tikzpicture}[scale=1]
\begin{axis}[legend pos=outer north east,
    title={Relative Pressure Jump Error},
    xlabel={Time, $s$},
    ylabel={$E(\Delta p)$},
]
\addplot [black] table [x={time}, y={ptotal1}] {\sta};
\addlegendentry{\small $h = 9.17 \cdot 10^{-3}$ $m$}
\addplot [red] table [x={time}, y={ptotal2}] {\sta};
\addlegendentry{\small $h = 8.4 \cdot 10^{-3}$ $m$}
\addplot [blue] table [x={time}, y={ptotal3}] {\sta};
\addlegendentry{\small $h = 7.75 \cdot 10^{-3}$ $m$}
\addplot [yellow] table [x={time}, y={ptotal4}] {\sta};
\addlegendentry{\small $h = 7.19 \cdot 10^{-3}$ $m$}
\addplot [green] table [x={time}, y={ptotal5}] {\sta};
\addlegendentry{\small $h = 6.71 \cdot 10^{-3}$ $m$}
\end{axis}
\path
    ([shift={( 5\pgflinewidth, 5\pgflinewidth)}]current bounding box.north east);
\end{tikzpicture}
\caption{Time evolution of the Relative Pressure Jump Error with different grid sizes. In all the tests $\tau = 10^{-5}$ $s$, $\sigma = 1$ $N/m$, $\rho_l = 10$ $kg/m^3$.}
\label{fig:21}
\end{figure}
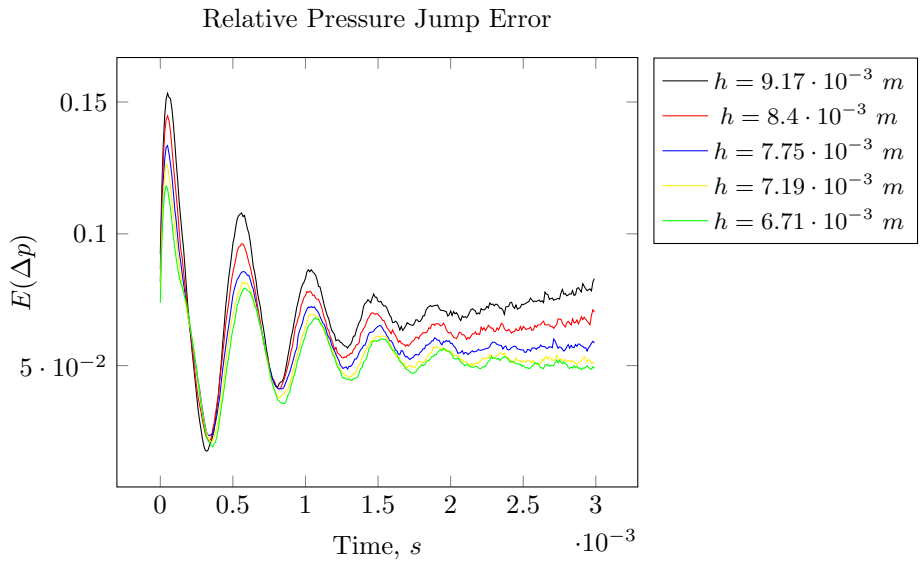

\begin{table}
\begin{tabular}{ |p{3cm}||p{3cm}|p{3cm}|  }
 \hline
 $\epsilon$& $E(\Delta p)$ & $||u||_{\infty}$\\
 \hline
 $0.5$   & $0.0755 $   &$1.79 \cdot 10^{-5}$\\
 $1$&   $0.0793$& $4.1 \cdot 10^{-4}$ \\
 $1.5$ &$0.1149$ & $2.4 \cdot 10^{-4}$\\
 \hline
\end{tabular}
\caption{$E(\Delta p)$ and $||u||_{\infty}$ for different values of $\epsilon$ (i.e. for different interface thickness).}
\label{tab:1}
\end{table}

\subsection{One-Dimensional Shock Wave-Interface Interaction.}
\label{Shock Wave-Interface Interaction. 1D.}
Next, a one-dimensional shock-interface interaction is studied using a test case from \cite{gfm} and \cite{wenocol}. This test verifies the performance of the GP-stabilization in the multicomponent case, as well as the robustness of the interface capturing approach. A strong (Mach 8.96) shockwave is travelling in helium ($\gamma = 1.667$, $\pi^{\infty} = 0$) towards a material interface with air ($\gamma = 1.4$, $\pi^{\infty} = 0$). Both materials are assumed to be inviscid. Following \cite{joh}, non-dimensional initial conditions are given by:

\begin{equation}
   (\rho,u,p,\phi) = 
   \begin{cases}  
  		(0.386,26.59,100,1) \text{~~~~~~~~~~~~~~if } 0 \leq x < 0.2\\
  		(0.1,-0.5,1,1) \text{~~~~~~~~~~~~~~~~~~~~~if } 0.2 \leq x < 0.8\\
  		(1,-0.5,1,0)  \text{~~~~~~~~~~~~~~~~~~~~~~~if } 0.8 \leq x < 2.\\
\end{cases}
 \end{equation}
 
Such problems are known to be challenging, since interface sharpening methods often lead to miscomputations of shock positions and speed (see \cite{wenocol}). Here, a numerical solution at $T_{end}=0.07$ ($\tau = 1 \cdot 10^{-5}$, $h=10^{-3}$) is compared with the exact one (derived in \cite{gfm}). The GP-dissipation term is used to stabilize the solution. The interface sharpening procedure is used with the parameter $\epsilon = 0.5$. The results are present in Figure \ref{fig:22}.

As can be seen, the interface remains very sharp, although density appears to be smeared in the regions adjacent to the artificial mixing layer. Shock positions and strengths are computed correctly and no oscillations are present, which validates both the interface capturing technique and the GP-stabilization term, as well as the general discretization strategy used in this study. So, the present method can be used to compute shock-interface interaction problems. 
\pgfplotstableread{swi.dat}{\swi}
\begin{figure}[htp]
  \begin{subfigure}[b]{.5\linewidth}
    \centering
\begin{tikzpicture}[scale=0.69]
\begin{axis}[legend pos= north west,
    title={Density},
    xlabel={x},
    ylabel={$\rho$}
]
\addplot [red] table [x={x}, y={rho1}] {\swi};
\addlegendentry{numerical}
\addplot [black,dashed] table [x={x}, y={rhoexact}] {\swi};
\addlegendentry{exact}
\end{axis}
 \path
    ([shift={( 5\pgflinewidth, 5\pgflinewidth)}]current bounding box.north east);
\end{tikzpicture}
    \subcaption{\label{fig:22a}}
  \end{subfigure}%
  \begin{subfigure}[b]{.5\linewidth}
    \centering
  \begin{tikzpicture}[scale=0.69]
\begin{axis}[legend pos=south west,
    title={Pressure},
    xlabel={x},
    ylabel={$p$}
]
\addplot [red] table [x={x}, y={p1}] {\swi};
\addlegendentry{numerical}
\addplot [black,dashed] table [x={x}, y={pexact}] {\swi};
\addlegendentry{exact}
\end{axis}
 \path
    ([shift={( 5\pgflinewidth, 5\pgflinewidth)}]current bounding box.north east);
\end{tikzpicture}
    \subcaption{\label{fig:22b}}
  \end{subfigure}
   \begin{subfigure}[b]{.5\linewidth}
    \centering
    \begin{tikzpicture}[scale=0.69]
\begin{axis}[legend pos=south west,
    title={Velocity},
    xlabel={x},
    ylabel={$u$}
]
\addplot [red] table [x={x}, y={u1}] {\swi};
\addlegendentry{numerical}
\addplot [black,dashed] table [x={x}, y={uex}] {\swi};
\addlegendentry{exact}
\end{axis}
 \path
    ([shift={( 5\pgflinewidth, 5\pgflinewidth)}]current bounding box.north east);
\end{tikzpicture}
    \subcaption{\label{fig:22c}}
  \end{subfigure}
     \begin{subfigure}[b]{.5\linewidth}
    \centering
    \begin{tikzpicture}[scale=0.69]
\begin{axis}[legend pos=south west,
    title={VoF-function},
    xlabel={x},
    ylabel={$\phi$}
]
\addplot [red] table [x={x}, y={phi1}] {\swi};
\addlegendentry{numerical}
\addplot [black,dashed] table [x={x}, y={phiexact}] {\swi};
\addlegendentry{exact}
\end{axis}
 \path
    ([shift={( 5\pgflinewidth, 5\pgflinewidth)}]current bounding box.north east);
\end{tikzpicture}
    \subcaption{\label{fig:22d}}
  \end{subfigure}
\caption{Density, pressure, velocity and VoF-function at $T_{end}=0.07$. Exact and numerical ($\tau = 1 \cdot 10^{-5}$, $h=10^{-3}$) solutions.}
\label{fig:22}
\end{figure}

\subsection{Shock Wave Refraction.}
\label{Shock Wave Refraction.}
Shock-interface interactions in two dimensions are known to produce nontrivial refraction patterns and they may serve as good tests for multidimensional, multicomponent algorithms at high Mach numbers. Following \cite{nourg}, a two-dimensional shock-interface interaction is studied using the Euler equations (no surface tension, viscous and heat transfer effects). In a square $[1m \times 1m]$ domain, a normal incident shock wave is propagating to the left in water ($\gamma = 4.4$, $\pi_{\infty} = 6 \cdot 10^8$ $Pa$) and impacting a planar material interface with air ($\gamma = 1.4$), inclined on an angle $\beta$.

\begin{figure}[htp]
  \begin{subfigure}[b]{.5\linewidth}
    \centering
    \includegraphics[width=2in]{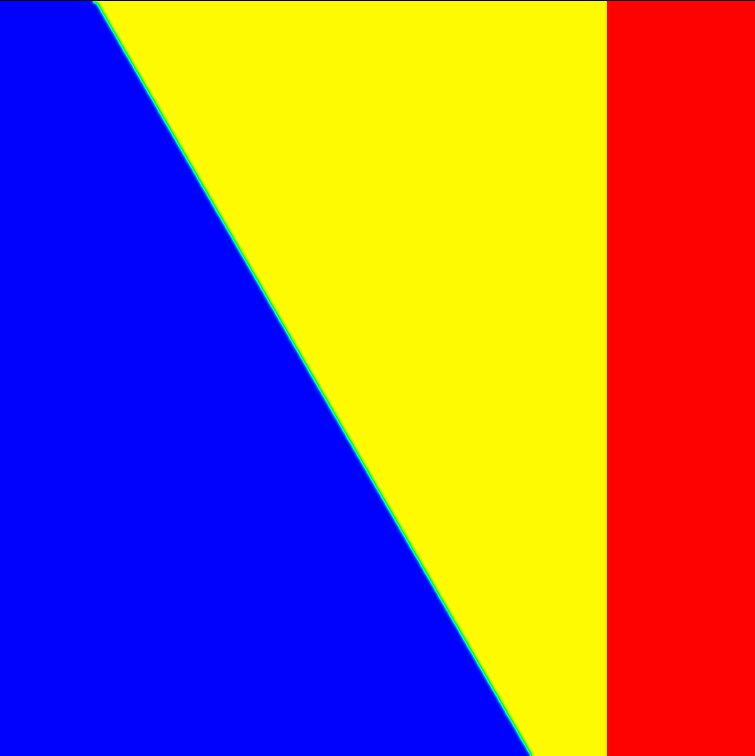}
    \subcaption{$\beta = \frac{\pi}{6}$}
    \label{fig:23a}
  \end{subfigure}%
  \begin{subfigure}[b]{.5\linewidth}
    \centering
    \includegraphics[width=2in]{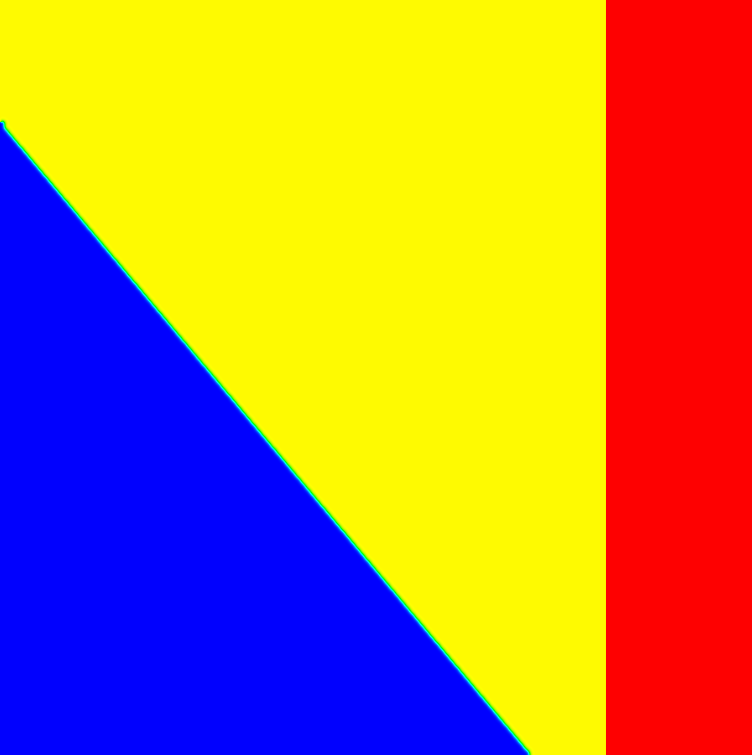}
    \subcaption{$\beta = \frac{\pi}{4.5}$}
    \label{fig:23b}
  \end{subfigure}
   \begin{subfigure}[b]{.5\linewidth}
    \centering
    \includegraphics[width=2in]{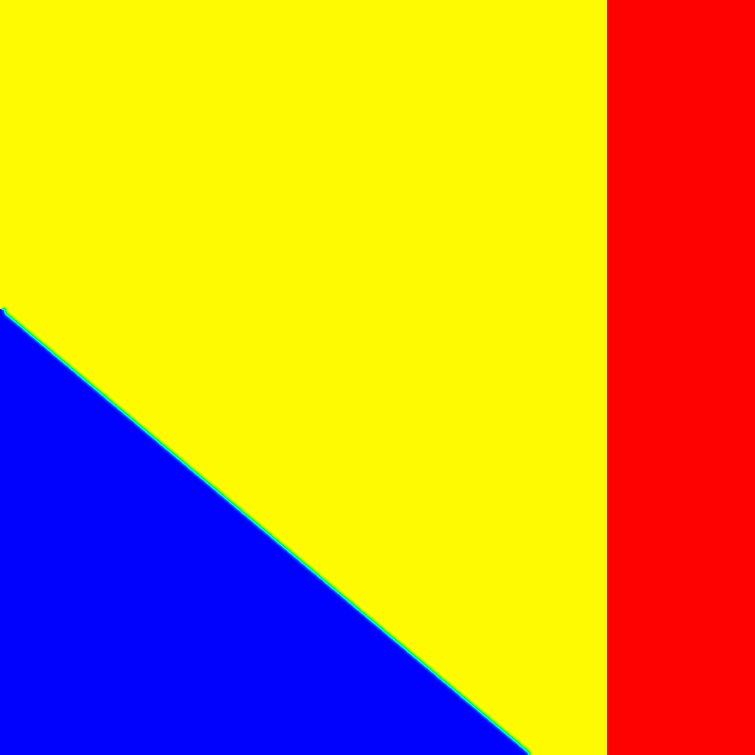}
    \subcaption{$\beta = \frac{\pi}{3.6}$}
    \label{fig:23c}
  \end{subfigure}
   \begin{subfigure}[b]{.5\linewidth}
    \centering
    \includegraphics[width=2in]{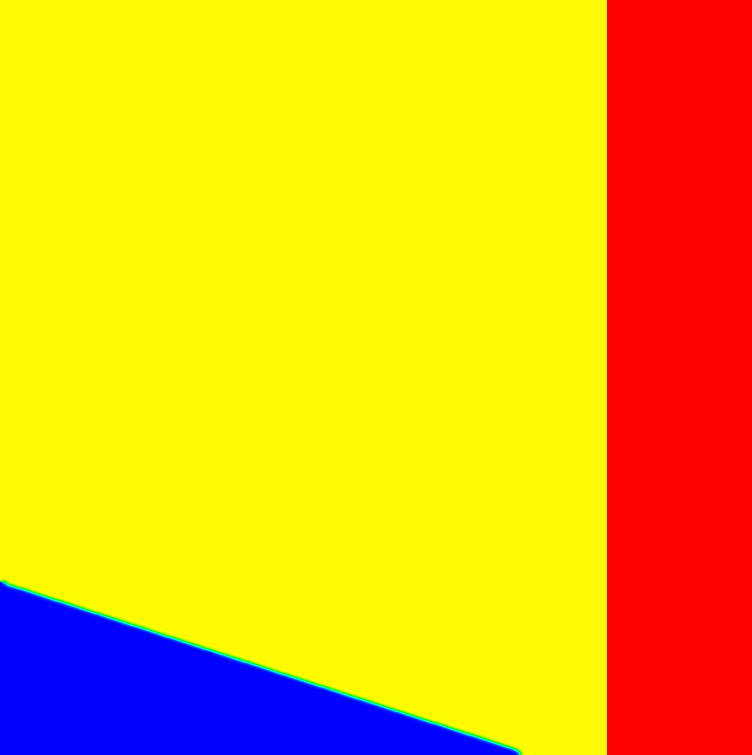}
    \subcaption{$\beta = \frac{\pi}{2.5}$}
    \label{fig:23d}
  \end{subfigure}
  \caption{Initial $\rho$-fields (pseudocolor) for different values of $\beta$. Red - water post-shock state, yellow - water pre-shock state, blue - air.}
  \label{fig:23}
\end{figure}

The following pre- and post-shock conditions are specified for water in terms of the primitive variables $(p,\rho,u,v)$ in $(Pa,kg/m^3,m/s,m/s)$ (similar to \cite{nourg}):
\begin{equation}
(p,\rho,u,v)_{pre} = (1 \cdot 10^5, 1000,0,0)
\end{equation}
\begin{equation}
(p,\rho,u,v)_{post} = (1.9 \cdot 10^9, 1323.65,-681.58,0),
\end{equation}
and the initial state of air is specified as:
\begin{equation}
(p,\rho,u,v)_{air} = (1 \cdot 10^5, 1,0,0).
\end{equation}
Initial fields of $\rho$ for angles $\beta = \frac{\pi}{6}$,$\frac{\pi}{4.5}$,$\frac{\pi}{3.6}$,$\frac{\pi}{2.5}$ are shown in Figure \ref{fig:23}.\\

Non-reflection boundary conditions (proposed in \cite{bcs}) are prescribed explicitly at the left, right, and top boundaries of the domain, and symmetry boundary conditions are used at the bottom. The computations are performed on a $800 \times 800$ uniform grid, with $\tau = 10^{-7}$ $s$. No splitting error reduction was used for this test.

\begin{figure}[htp]
  \begin{subfigure}[b]{.5\linewidth}
    \centering
    \includegraphics[width=2in]{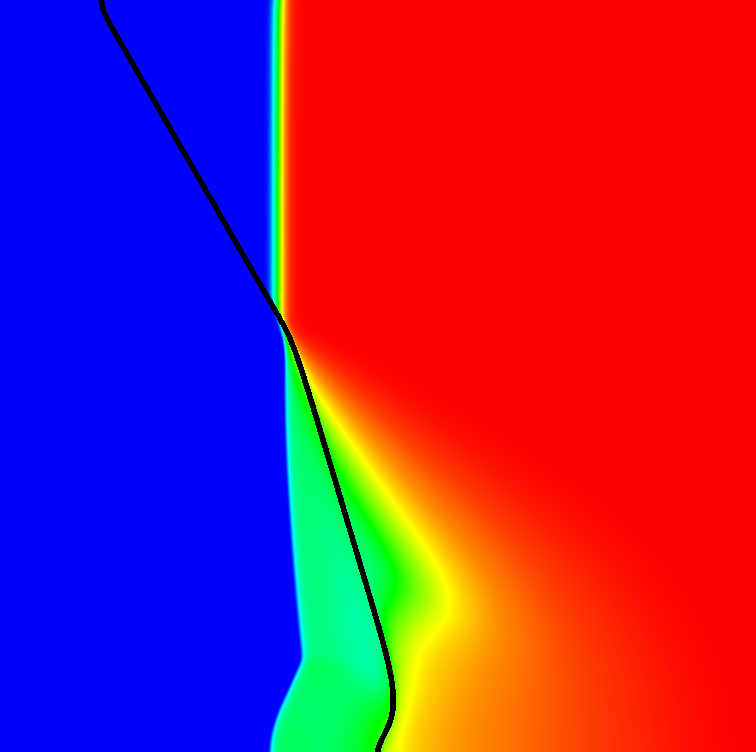}
    \caption{$\log{p}$ and $\phi$}
    \label{fig:24a}
  \end{subfigure}%
  \begin{subfigure}[b]{.5\linewidth}
    \centering
    \includegraphics[width=2in]{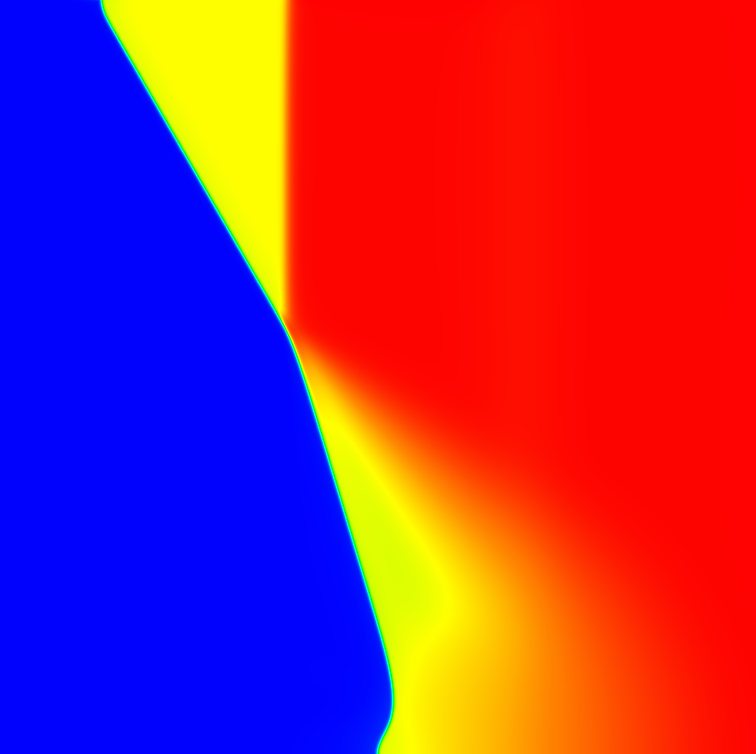}
    \subcaption{$\rho$}
    \label{fig:24b}
  \end{subfigure}
    \caption{$\log{p}$ (pseudocolor), $\phi$ (contour), and $\rho$ (pseudocolor) plots at $T_{end} = 1.5 \cdot 10^{-4}$ $s$ for $\beta = \frac{\pi}{6}$}
  \label{fig:24}
\end{figure}

\begin{figure}[htp]
   \begin{subfigure}[b]{.5\linewidth}
    \centering
    \includegraphics[width=2in]{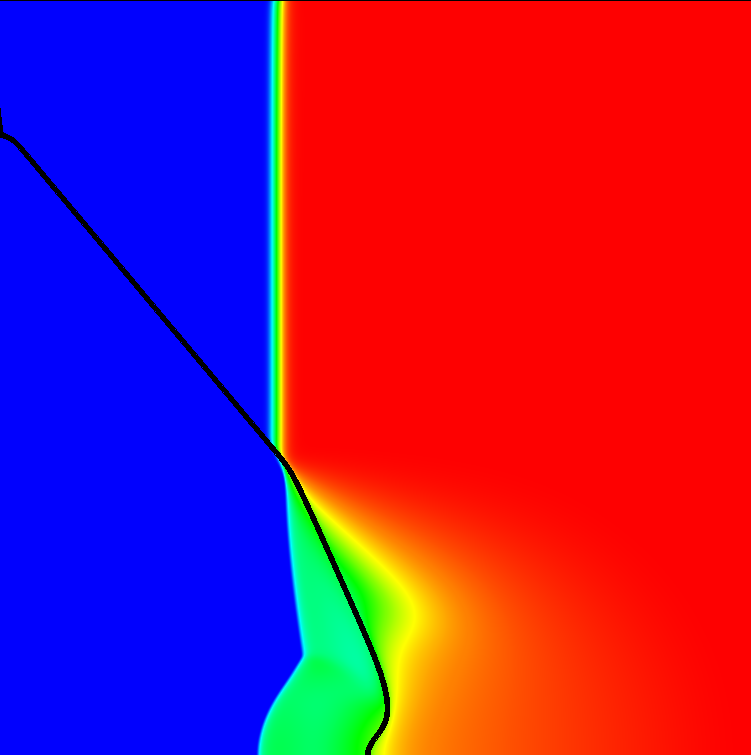}
    \subcaption{$\log{p}$ and $\phi$}
    \label{fig:25a}
  \end{subfigure}
   \begin{subfigure}[b]{.5\linewidth}
    \centering
    \includegraphics[width=2in]{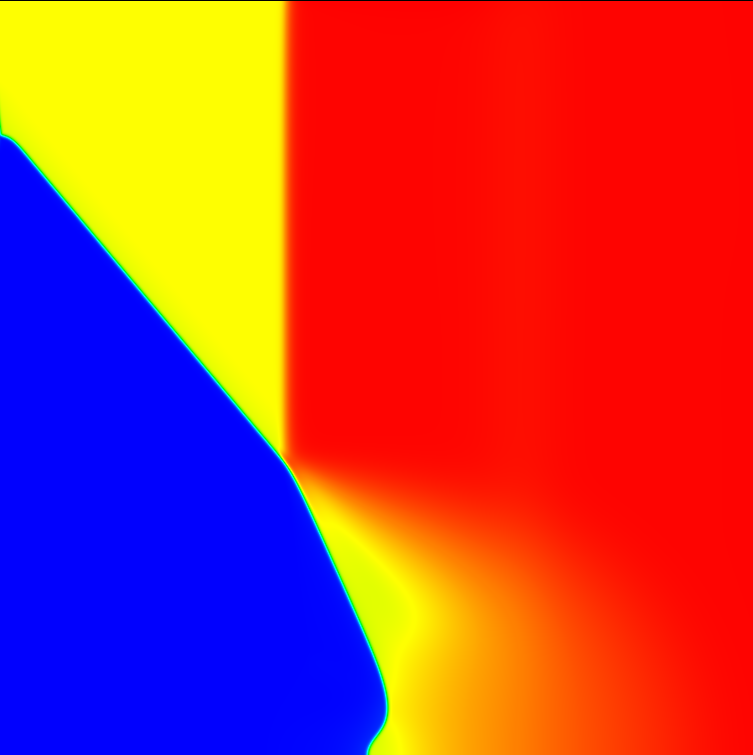}
    \subcaption{$\rho$}
    \label{fig25b}
  \end{subfigure}
    \caption{$\log{p}$ (pseudcolor), $\phi$ (contour), and $\rho$ (pseudocolor) plots at $T_{end} = 1.5 \cdot 10^{-4}$ $s$ for $\beta = \frac{\pi}{4.5}$}
  \label{fig:25}
\end{figure}

\begin{figure}[htp]
  \begin{subfigure}[b]{.5\linewidth}
    \centering
    \includegraphics[width=2in]{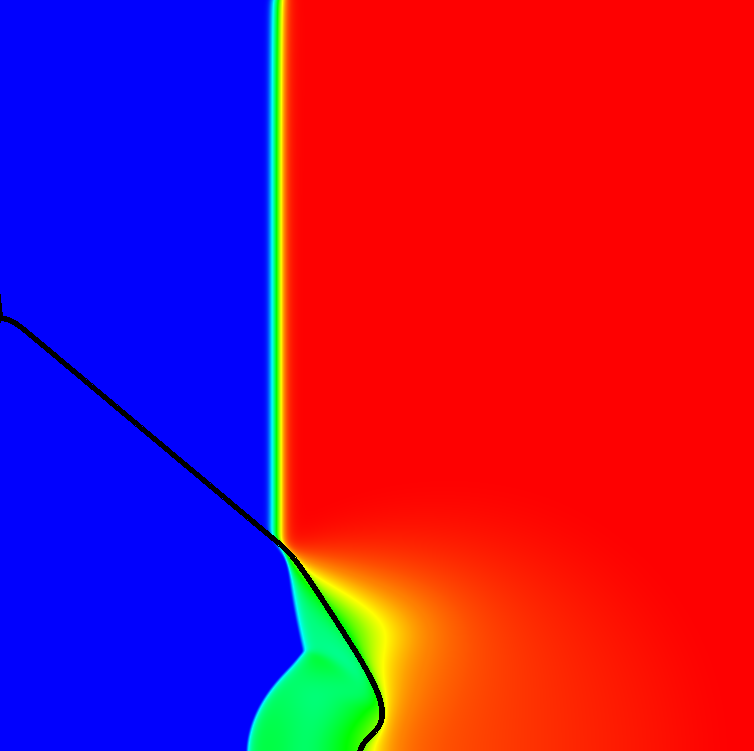}
    \subcaption{$\log{p}$ and $\phi$}
    \label{fig:26a}
  \end{subfigure}
   \begin{subfigure}[b]{.5\linewidth}
    \centering
    \includegraphics[width=2in]{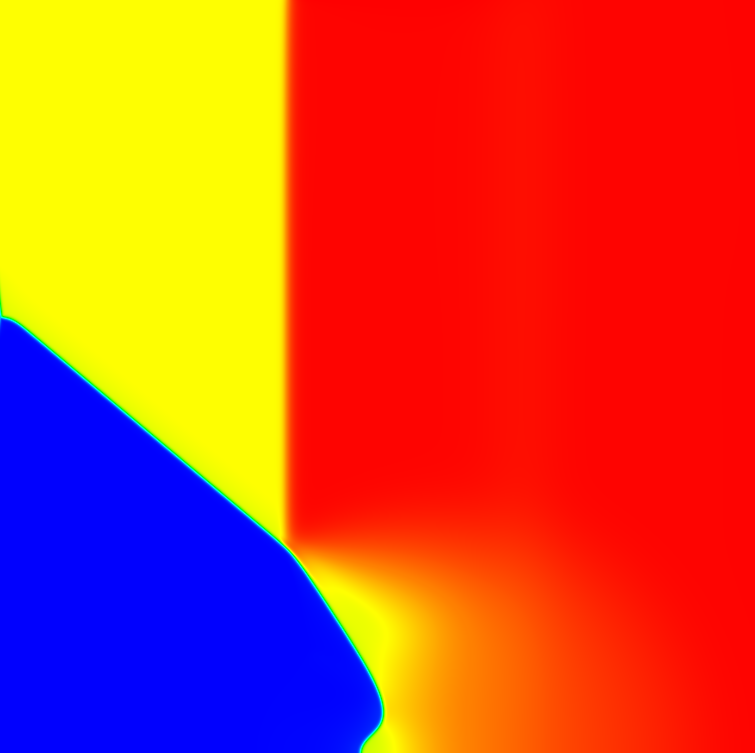}
    \subcaption{$\rho$}
    \label{fig:26b}
  \end{subfigure}
   \caption{$\log{p}$ (pseudcolor), $\phi$ (contour), and $\rho$ (pseudocolor) plots at $T_{end} = 1.5 \cdot 10^{-4}$ $s$ for $\beta = \frac{\pi}{3.6}$}
  \label{fig:26}
\end{figure}

\begin{figure}[htp]
  \begin{subfigure}[b]{.5\linewidth}
    \centering
    \includegraphics[width=2in]{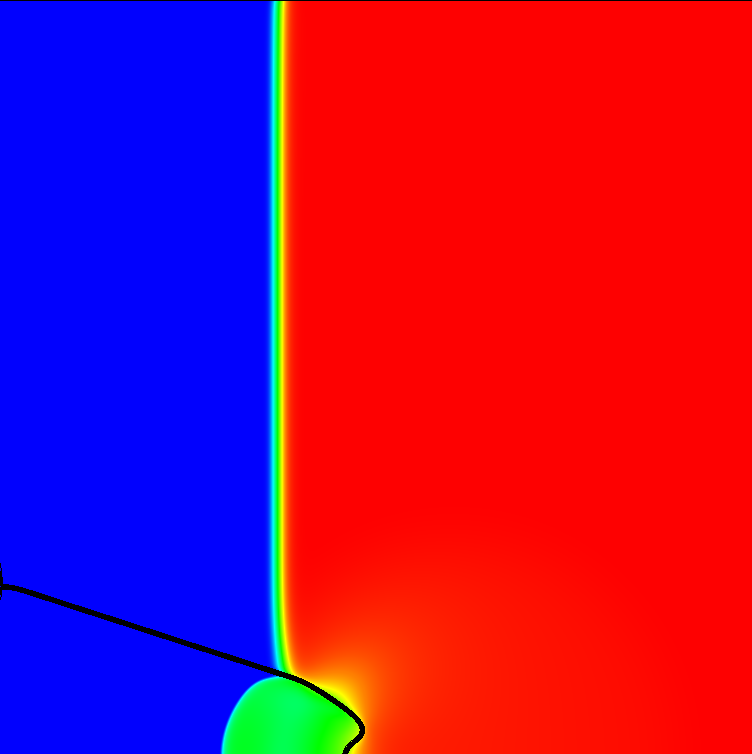}
    \subcaption{$\log{p}$ and $\phi$}
    \label{fig:27a}
  \end{subfigure}
   \begin{subfigure}[b]{.5\linewidth}
    \centering
    \includegraphics[width=2in]{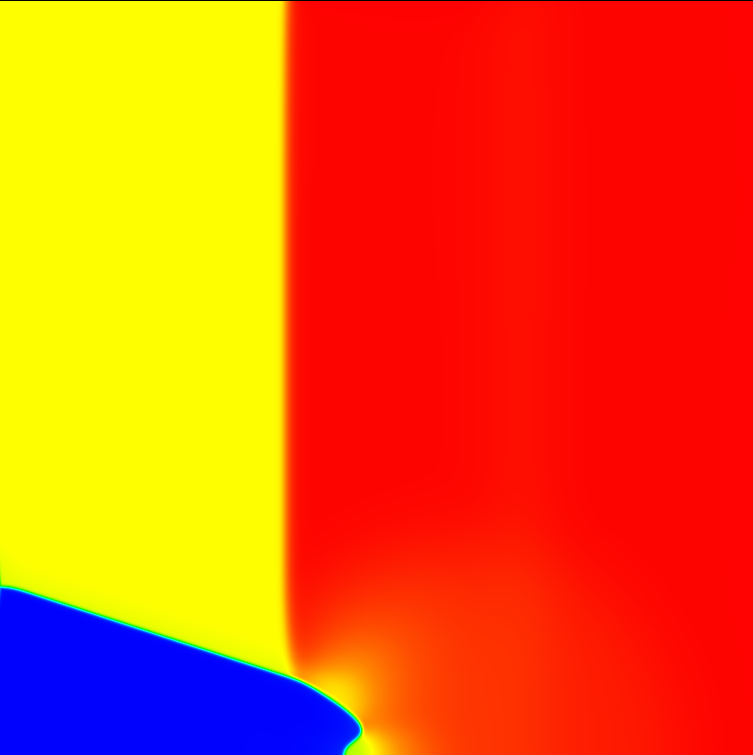}
    \subcaption{$\rho$}
    \label{fig:27b}
  \end{subfigure}
  \caption{$\log{p}$ (pseudcolor), $\phi$ (contour), and $\rho$ (pseudocolor) plots at $T_{end} = 1.5 \cdot 10^{-4}$ $s$ for $\beta = \frac{\pi}{2.5}$}
  \label{fig:27}
\end{figure}

Figures \ref{fig:24}-\ref{fig:27} represent $\log{p}$, $\phi$, and $\rho$ at $T_{end} = 1.5 \cdot 10^{-4}$ $s$. The results are in good agreement with those from \cite{nourg}. The material interface remains thin throughout the computations ($\epsilon = 1$), and the refraction patterns are captured correctly. The cases of $\beta = \frac{\pi}{6}$,$\frac{\pi}{4.5}$,$\frac{\pi}{3.6}$ produce regular refraction patterns with reflected expansion (RRE), while the case of $\beta = \frac{\pi}{2.5}$ gives rise to irregular concave forward refraction (CFR), as the impacted wave interacts with the reflected expansion, ``which results in a mutual annihilation and formation of a compound wave" (see \cite{nourg}). The compound wave is curved due to the speed difference between the incident shock and the refraction node (intersection of the compound and transmitted waves), and weaker than the impacting shock. In all the test cases a water jet is formed by the convergence of shock-induced flows, which is also in agreement with the results of \cite{nourg}. Therefore, the algorithm has demonstrated its ability to simulate multidimensional inviscid flows featuring shock-interface interactions without topological changes and successfully capture nontrivial refraction patterns.


\subsection{Shock Wave-Bubble Interaction.}
\label{Shock Wave-Bubble Interaction.}
Finally, a shock-curved interface interaction test is considered to validate the applicability of the method to this class of problems and demonstrate the possibility of handling topological changes. In this test case the same materials and the same initial conditions (pre-, post-shock states, location of the shock wave) are used as in the previous test, but the interface forms an air bubble of radius $R=0.2m$ centered at $(0.55m,0.5m)$. The full system of NSEs is considered here, with $\mu_{water} = 10^{-3}$ $kg/(s \cdot m)$, $\mu_{air} = 10^{-5}$ $kg/(s \cdot m)$, $\sigma = 0.073$ $N/m$. Interface sharpening is performed with $\epsilon = 1.5$. Initial conditions for air are given by:
\begin{equation}
(p,\rho,u,v)_{air} = (1 \cdot 10^5 + \frac{\sigma}{R}, 1,0,0).
\end{equation}

\begin{figure}[htp]
  \begin{subfigure}[b]{.5\linewidth}
    \centering
    \includegraphics[width=2in]{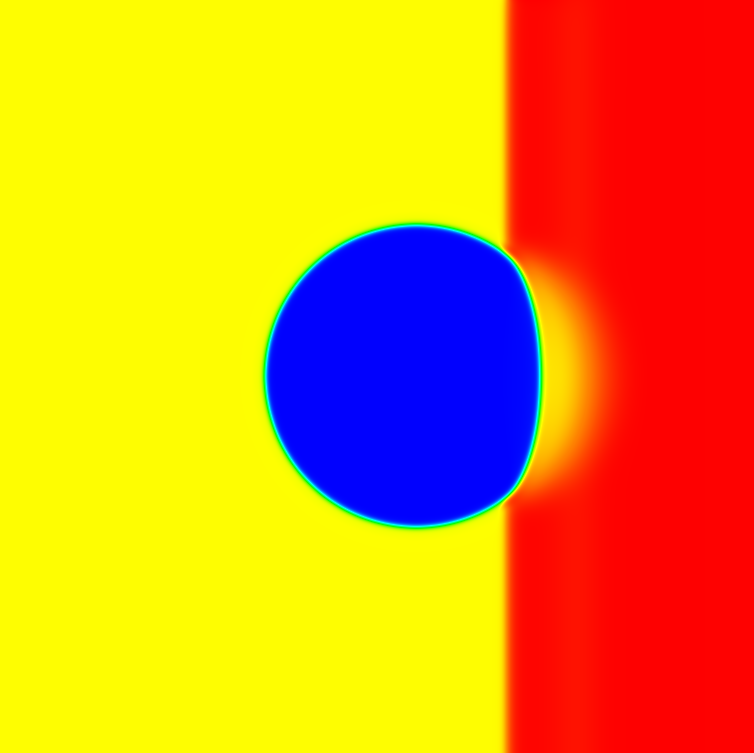}
    \subcaption{$\rho$}
    \label{fig:28a}
  \end{subfigure}
  \begin{subfigure}[b]{.5\linewidth}
    \centering
    \includegraphics[width=2in]{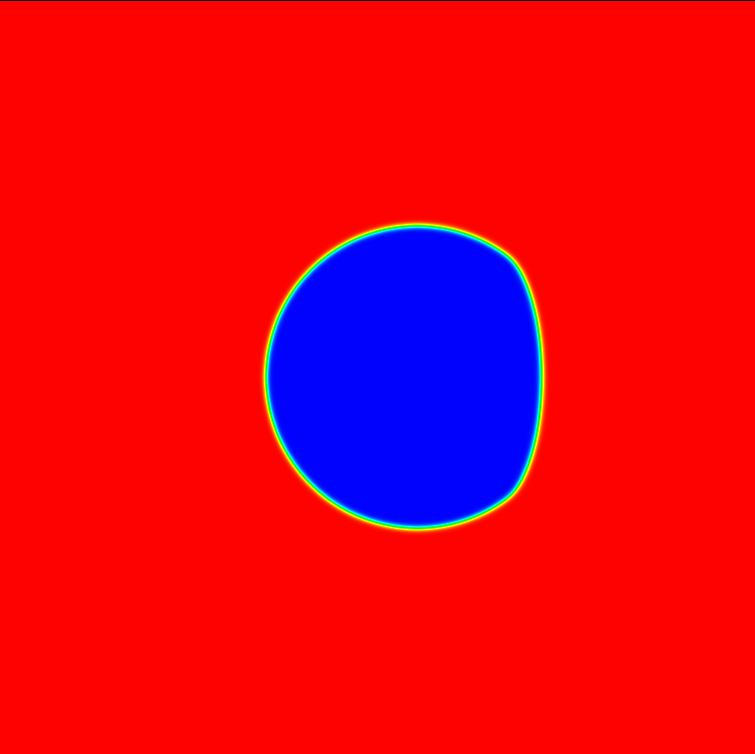}
    \subcaption{$\phi$}
    \label{fig:28b}
  \end{subfigure}
    \caption{$\rho$ and $\phi$ pseudocolor fields at $t = 4.5 \cdot 10^{-5}$ $s$}
  \label{fig:28}
\end{figure}

\begin{figure}[htp]
  \begin{subfigure}[b]{.5\linewidth}
    \centering
    \includegraphics[width=2in]{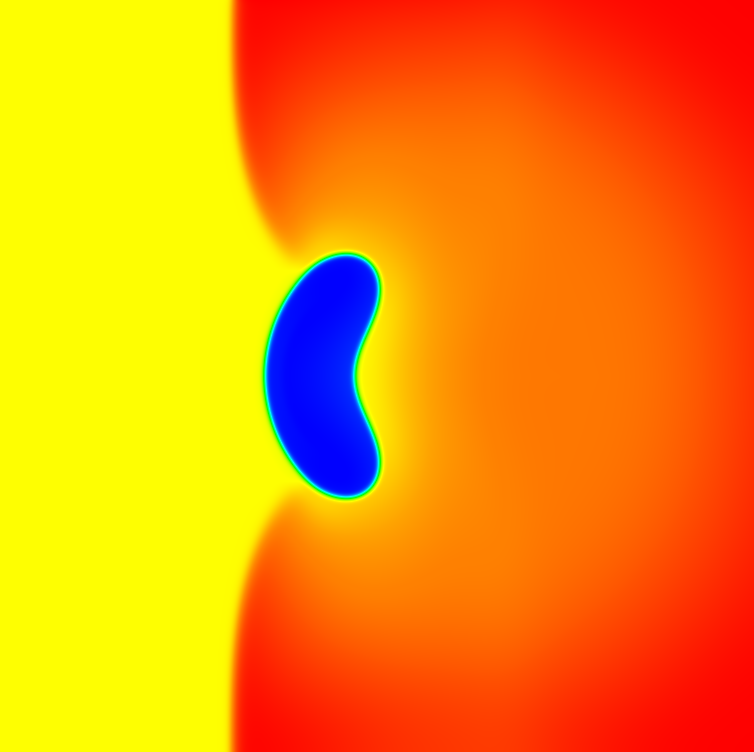}
    \subcaption{$\rho$}
    \label{fig:31a}
  \end{subfigure}
   \begin{subfigure}[b]{.5\linewidth}
    \centering
    \includegraphics[width=2in]{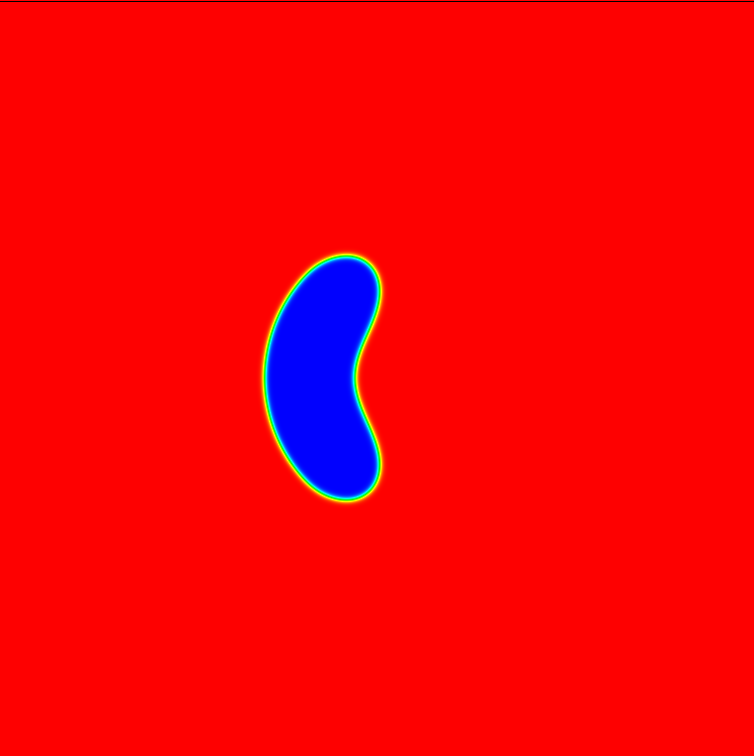}
    \subcaption{$\phi$}
    \label{fig:31b}
  \end{subfigure}
    \caption{$\rho$ and $\phi$ pseudocolor fields at $t = 1.75 \cdot 10^{-4}$ $s$}
  \label{fig:31}
\end{figure}

\begin{figure}[htp]
   \begin{subfigure}[b]{.5\linewidth}
    \centering
    \includegraphics[width=2in]{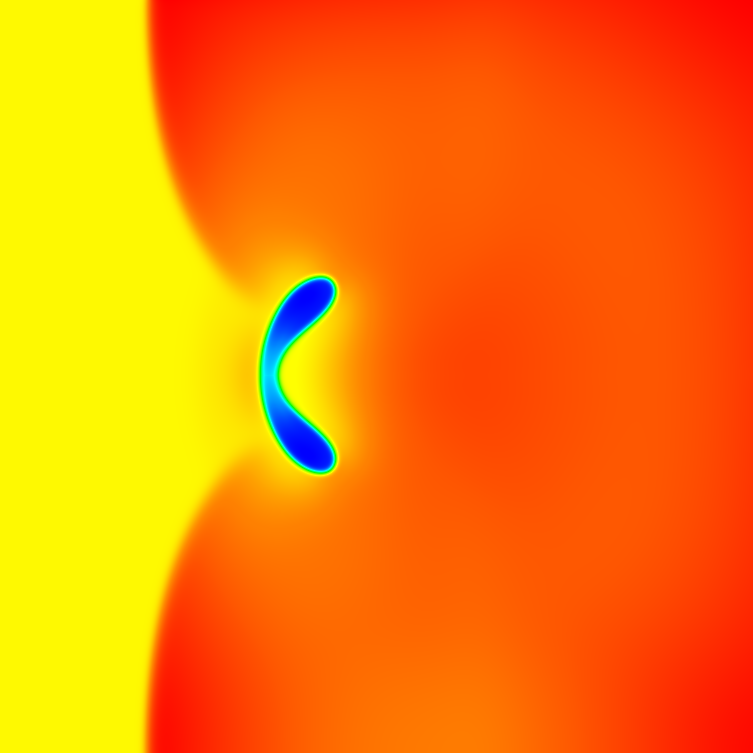}
    \subcaption{$\rho$}
    \label{fig:32a}
  \end{subfigure}
   \begin{subfigure}[b]{.5\linewidth}
    \centering
    \includegraphics[width=2in]{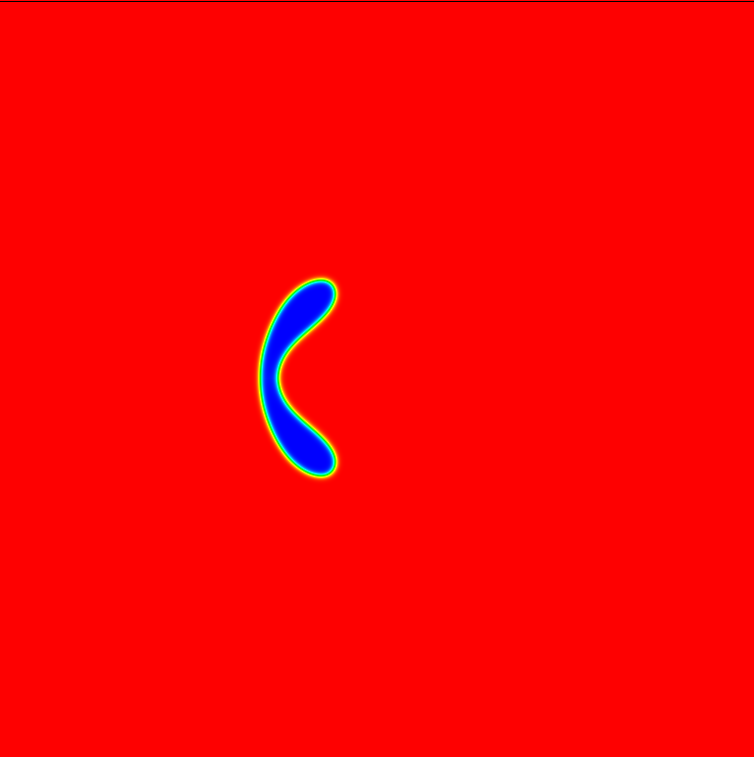}
    \subcaption{$\phi$}
    \label{fig:32b}
  \end{subfigure}
    \caption{$\rho$ and $\phi$ pseudocolor fields at $t = 2.16 \cdot 10^{-4}$ $s$}
  \label{fig:32}
\end{figure}

\begin{figure}[htp]
   \begin{subfigure}[b]{.5\linewidth}
    \centering
    \includegraphics[width=2in]{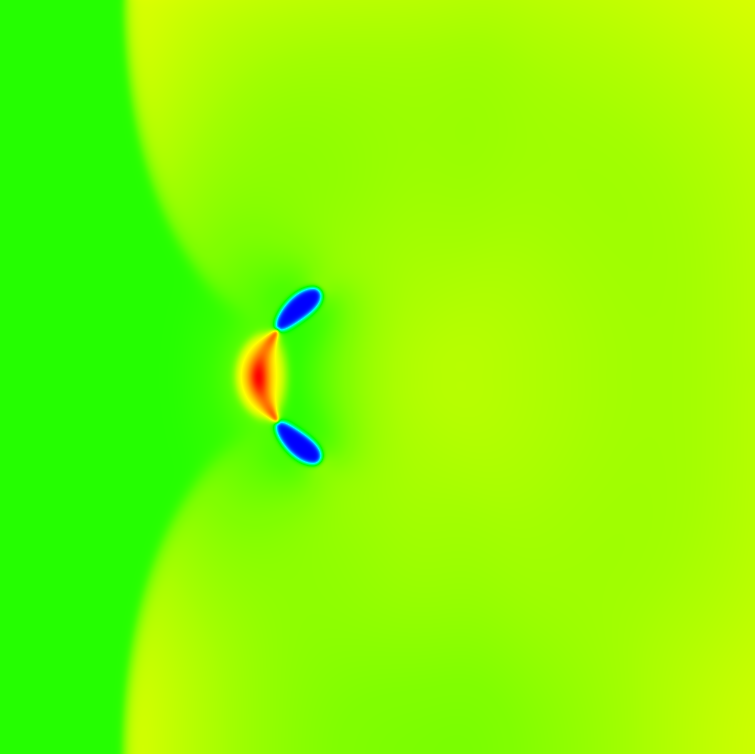}
    \subcaption{$\rho$}
    \label{fig:34a}
  \end{subfigure}
   \begin{subfigure}[b]{.5\linewidth}
    \centering
    \includegraphics[width=2in]{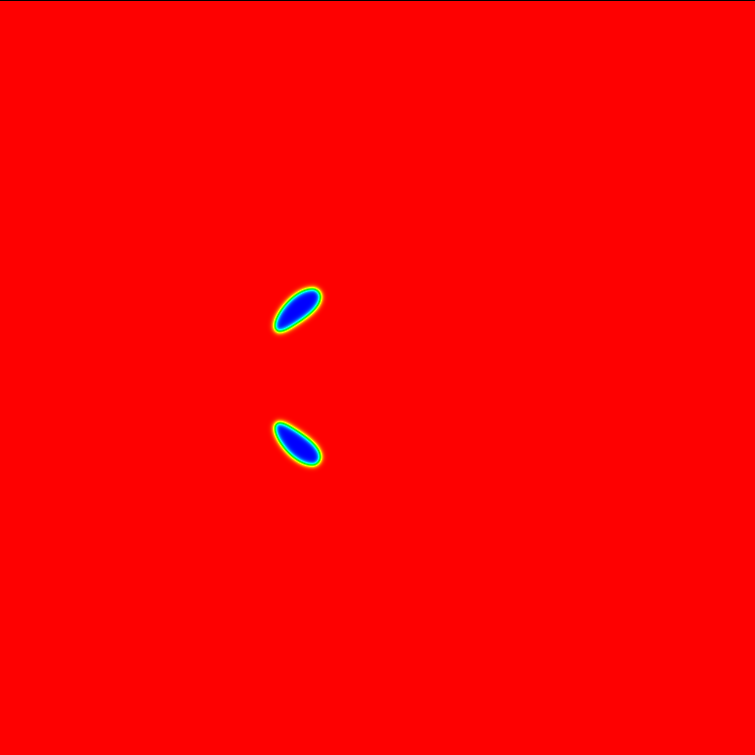}
    \subcaption{$\phi$}
    \label{fig:34b}
  \end{subfigure}
 \caption{$\rho$ and $\phi$ pseudocolor fields at $t = 2.276 \cdot 10^{-4}$ $s$}
  \label{fig:34}
\end{figure}

The computations are performed on a $800 \times 800$ uniform grid, with $\tau = 10^{-7}$ $s$. No splitting error reduction was used for this test. The fields $\rho$ and $\phi$ are presented at different times in Figures \ref{fig:28}-\ref{fig:34}. The results are in good agreement with \cite{sharp}, where a similar test case was presented. The interface remains thin throughout the computations, and the effects of wave refraction and eventual topological change (bubble collapse) with the formation of a high-speed jet are well-captured. This test reveals the potential of the scheme to solve a full system of multicomponent NSEs in the high-Mach number regime, as well as the ability of the methods to handle topological changes of the interface.\\


  \clearpage
\section{Conclusions.}
\label{Conclusion.}
In this paper, an efficient numerical algorithm applicable to a wide range of compressible multicomponent flows has been built based on the linearized block implicit (LBI) factored schemes. The main contribution of this study to the development of such schemes is the splitting error reduction technique, which enables their use for low-Mach number flows.

An artificial dissipation term for high-Mach number applications has been constructed as a finite difference interpretation of a novel finite element technique proposed by Guermond and Popov from \cite{invdom}, which introduces artificial viscosity coefficients based on the maximum speed of propagation in local one-dimensional Riemann problems. The estimation algorithm for the maximum speed of propagation proposed in \cite{maxspeed} has been extended here to the case of the stiffened gas equation of state. The results presented in this paper show the advantages of the approach proposed in \cite{invdom}, and continue the development of this class of schemes.

Another contribution of the present paper is the novel artificial dissipation term designed here for low-Mach number applications. It is based on second-order differences of conservative variables and products of conservative variables and their corresponding Jacobians. Such a combination is shown to be effective in eliminating the odd-even decoupling problem while controlling the total kinetic energy of the system. This allows for the use of LBI factored schemes in terms of conservative variables on non-staggered grids for low-Mach number applications.

A consistent coupling of this technique with the interface-capturing approach without disrupting the interfacial equilibrium is achieved by introducing functions of EOS coefficients as variables, as well as the special design of the stabilization terms. The sharpening technique from \cite{sharp} is shown to be compatible with the method, and with the CSF implementation of the surface tension from \cite{sten}.

Possible extensions and directions for further research may include the use of more accurate time-marching methods or a defect-correction technique to lift the order of accuracy of the algorithm, incorporating implicit non-reflection boundary conditions, and extending the method to the three-dimensional case and more complicated geometries.

\section{Acknowledgements.}
This work was supported by a Discovery Grant \#219949 of the Natural Sciences and
Engineering Research Council of Canada.  Acknowledgment is also made to the donors of the American Chemical Society Petroleum Research Fund for partial support of this research,  by a grant \# 55484-ND9.

I wish to thank professor R. Krechetnikov for suggesting the original idea and problem formulation for the project, professor P. Minev for his guidance and pieces of advice throughout this work, the creators of the VisIt (see \cite{VisIt}) and ParaView (see \cite{paraview}) applications and the LAPACK (see \cite{lapack}) library for providing extremely helpful pieces of software, and the Compute Canada (www.computecanada.ca) organization for providing computing hardware and technical support. 

\section{Appendix A. Governing Equations and computation of Jacobians.} \label{Appendix A}
The governing equations (compressible Navier-Stokes and advection of SG EOS coefficients) can be written as:\\
\begin{equation}
\label{B1}
\begin{split}
\frac{\partial U}{ \partial t} + A\frac{\partial F(U)}{\partial x} + B\frac{\partial G(U)}{\partial y} = 
\frac{\partial V_1 (U,U_x)}{\partial x} &+ \frac{\partial V_2 (U,U_y)}{\partial x} \\+ \frac{\partial W_1(U,U_x)}{\partial y} + \frac{\partial W_2(U,U_y)}{\partial y} + F^{ST},
\end{split} 
\end{equation}
where
\begin{equation}
U = 
\begin{bmatrix}
\rho&m & n & E & \alpha & \beta
\end{bmatrix}
^T
\end{equation}
\begin{equation}
F = 
\begin{bmatrix}
m & \frac{m^2}{\rho} + p & \frac{mn}{\rho} & (E+p)\frac{m}{\rho} & \alpha & \beta
\end{bmatrix}
^T
\end{equation}
\begin{equation}
G = 
\begin{bmatrix}
n & \frac{nm}{\rho} & \frac{n^2}{\rho} + p & (E+p)\frac{n}{\rho} & \alpha & \beta
\end{bmatrix}
^T
\end{equation}
\begin{equation}
F^{ST} =\begin{bmatrix}
0 & \sigma \kappa \partial_x \phi & \sigma \kappa \partial_y \phi & \sigma \kappa u \partial_x \phi + \sigma \kappa v \partial_y \phi & 0 & 0
\end{bmatrix}
^T
\end{equation}
\begin{equation}
V_1 = 
\begin{bmatrix}
0 & \frac{4 \mu}{3}\partial_x \frac{m}{\rho} & \mu \partial_x \frac{n}{\rho} & \frac{4 \mu m}{3 \rho} \partial_x \frac{m}{\rho} + \mu \frac{n}{\rho} \partial_x \frac{n}{\rho} & 0 & 0
\end{bmatrix}
^T
\end{equation}
\begin{equation}
V_2 = 
\begin{bmatrix}
0 & \frac{-2 \mu}{3}\partial_y \frac{n}{\rho} & \mu \partial_y \frac{m}{\rho} & \frac{-2 \mu m}{3 \rho} \partial_y \frac{n}{\rho} + \mu \frac{n}{\rho} \partial_y \frac{m}{\rho} & 0 & 0
\end{bmatrix}
^T
\end{equation}
\begin{equation}
W_1 = 
\begin{bmatrix}
0 & \mu \partial_x \frac{n}{\rho} & -\frac{2 \mu}{3} \partial_x \frac{m}{\rho} & \frac{-2 \mu n}{3 \rho} \partial_x \frac{m}{\rho} + \mu \frac{m}{\rho} \partial_x \frac{n}{\rho} & 0 & 0
\end{bmatrix}
^T
\end{equation}
\begin{equation}
W_2 = 
\begin{bmatrix}
0 & \mu \partial_y \frac{m}{\rho} & \frac{4 \mu}{3} \partial_y \frac{n}{\rho} & \frac{ \mu m}{\rho} \partial_y \frac{m}{\rho} +  \frac{4 \mu n}{3 \rho} \partial_y \frac{n}{\rho} & 0 & 0
\end{bmatrix}
^T
\end{equation}\\
where $\mu$ $[kg/(s \cdot m)]$ is a dynamic viscosity coefficient, $\sigma$ $[N/m]$ is the surface tension coeficient, $\phi$ is the volume of fluid function, $\kappa = - \nabla \cdot \frac{\nabla \phi}{|\nabla \phi|}$ is the interfacial curvature.\\
\begin{equation}
A = 
\begin{bmatrix}
1 & 0 & 0 & 0 & 0 & 0 \\
0 & 1 & 0 & 0 & 0 & 0 \\
0 & 0 & 1 & 0 & 0 & 0 \\
0 & 0 & 0 & 1 & 0 & 0 \\
0 & 0 & 0 & 0 & \frac{m}{\rho} & 0 \\
0 & 0 & 0 & 0 & 0 & \frac{m}{\rho} \\
\end{bmatrix}
\end{equation}
\begin{equation}
B = 
\begin{bmatrix}
1 & 0 & 0 & 0 & 0 & 0 \\
0 & 1 & 0 & 0 & 0 & 0 \\
0 & 0 & 1 & 0 & 0 & 0 \\
0 & 0 & 0 & 1 & 0 & 0 \\
0 & 0 & 0 & 0 & \frac{n}{\rho} & 0 \\
0 & 0 & 0 & 0 & 0 & \frac{n}{\rho} \\
\end{bmatrix}
\end{equation}\\
\begin{equation}
p = \frac{1}{\alpha} \left ( E - \frac{m^2 + n^2}{2 \rho}  - \beta \right ).
\end{equation}\\
After time discretization (\ref{B1}) becomes:\\
\begin{equation}
\begin{split}
\frac{U^{n+1} - U^n}{\tau} + & A^n \frac{\partial F^{n+1}(U)}{\partial x} +  B^n\frac{\partial G^{n+1}(U)}{\partial y} =\\  \frac{\partial V_1^{n+1}(U,U_x)}{\partial x} &+ \frac{\partial V_2^{n}(U,U_y)}{\partial x} + \frac{\partial W_1^{n}(U,U_x)}{\partial y} + \frac{\partial W_2^{n+1}(U,U_y)}{\partial y},
\end{split} 
\end{equation} \\
which is linearized as:\\
\begin{equation}
F^{n+1} = F^n + \left ( \frac{\partial F}{\partial U} \right )^n (U^{n+1} - U^n)
\end{equation}
\begin{equation}
G^{n+1} = G^n + \left ( \frac{\partial G}{\partial U} \right )^n (U^{n+1} - U^n)
\end{equation}
\begin{equation}
V_1^{n+1} = V_1^n + \left ( \frac{\partial V_1}{\partial U} \right )^n (U^{n+1} - U^n) + \left ( \frac{\partial V_1}{\partial U_x} \right )^n (U_x^{n+1} - U_x^n)
\end{equation}
\begin{equation}
W_2^{n+1} = W_2^n + \left ( \frac{\partial W_2}{\partial U} \right )^n (U^{n+1} - U^n) + \left ( \frac{\partial W_2}{\partial U_y} \right )^n (U_y^{n+1} - U_y^n),
\end{equation}\\
where\\
\begin{equation}
\frac{\partial F}{\partial U} = 
\begin{bmatrix}
0 & 1 & 0 & 0 & 0 & 0 \\
f_1 & \frac{2 m}{\rho} - \frac{m}{\alpha \rho} &  - \frac{n}{\alpha \rho} & \frac{1}{\alpha} & f_2 & -\frac{1}{\alpha} \\
-\frac{mn}{\rho^2} & \frac{n}{\rho} & \frac{m}{\rho} & 0 & 0 & 0 \\
f_3 & \frac{E+p}{\rho} - \frac{m^2}{\alpha \rho^2} & - \frac{mn}{\alpha \rho^2} & \frac{m}{\rho} \left ( 1 + \frac{1}{\alpha}  \right ) & f_4 & -\frac{m}{\alpha \rho} \\
0 & 0 & 0 & 0 & 1 & 0 \\
0 & 0 & 0 & 0 & 0 & 1 \\
\end{bmatrix}
\end{equation}\\
where $f_1 = -\frac{m^2}{\rho^2} + \frac{1}{\alpha} \left ( \frac{m^2 + n^2}{2 \rho^2}  \right )$, $f_2 = -\frac{1}{\alpha^2} \left (E - \frac{m^2 + n^2}{2 \rho} -\beta  \right )$,\\ $f_3 = -\frac{(E+p)m}{\rho^2} + \frac{m^3 + mn^2}{2 \alpha \rho^3 }$, $f_4 = -\frac{m}{\alpha^2 \rho} \left ( E - \frac{m^2 + n^2}{2 \rho} -\beta  \right )$.\\
\begin{equation}
\frac{\partial G}{\partial U}= 
\begin{bmatrix}
0 & 0 & 1 & 0 & 0 & 0 \\
-\frac{mn}{\rho^2} & \frac{n}{\rho} & \frac{m}{\rho} & 0 & 0 & 0 \\
 g_1 &  - \frac{m}{\alpha \rho}& \frac{2 n}{\rho} - \frac{n}{\alpha \rho}  & \frac{1}{\alpha} & g_2 & -\frac{1}{\alpha} \\
g_3 &  - \frac{mn}{\alpha \rho^2} &\frac{E+p}{\rho} - \frac{n^2}{\alpha \rho^2} & \frac{n}{\rho} \left ( 1 + \frac{1}{\alpha}  \right ) & g_4 & -\frac{n}{\alpha \rho} \\
0 & 0 & 0 & 0 & 1 & 0 \\
0 & 0 & 0 & 0 & 0 & 1 \\
\end{bmatrix}
\end{equation}\\
where $g_1 = -\frac{n^2}{\rho^2} + \frac{1}{\alpha} \left ( \frac{m^2 + n^2}{2 \rho^2}  \right )$, $g_2 = -\frac{1}{\alpha^2} \left (E - \frac{m^2 + n^2}{2 \rho} -\beta  \right )$,\\ $g_3 = -\frac{(E+p)n}{\rho^2} + \frac{n^3 + nm^2}{2 \alpha \rho^3 }$, $g_4 = -\frac{n}{\alpha^2 \rho} \left ( E - \frac{m^2 + n^2}{2 \rho} -\beta  \right )$.\\
\begin{equation}
\frac{\partial V_1}{\partial U} = 
\begin{bmatrix}
0 & 0 & 0 & 0 & 0 & 0 \\
-\frac{4 \mu}{3} \left ( \frac{\partial_x m}{\rho^2} - \frac{2 m \partial_x \rho}{\rho^3} \right ) & -\frac{4 \mu}{3} \frac{\partial_x \rho}{\rho^2} & 0 & 0 & 0 & 0 \\
-\mu \left ( \frac{\partial_x n}{\rho^2} - \frac{2 n \partial_x \rho}{\rho^3} \right ) & 0 & -\mu \frac{\partial_x \rho}{\rho^2} & 0 & 0 & 0 \\
v_1 & v_2 & v_3 & 0 & 0 & 0 \\
0 & 0 & 0 & 0 & 0 & 0 \\
0 & 0 & 0 & 0 & 0 & 0 \\
\end{bmatrix}
\end{equation}\\
where $v_1 = -\frac{4 \mu}{3} \left (\frac{2 m \partial_x m}{\rho^3} - \frac{3 m^2 \partial_x \rho}{\rho^4}  \right ) - \mu \left (\frac{2 n \partial_x n}{\rho^3} - \frac{3 n^2 \partial_x \rho}{\rho^4}  \right )$, $v_2 = \frac{4 \mu}{3} \left ( \frac{\partial_x m}{\rho^2} - \frac{2 m \partial_x \rho}{\rho^3}  \right )$, $v_3 = \mu \left ( \frac{\partial_x n}{\rho^2} - \frac{2 n \partial_x \rho}{\rho^3}  \right )$.\\
\begin{equation}
\frac{\partial V_1}{\partial U_x} = 
\begin{bmatrix}
0 & 0 & 0 & 0 & 0 & 0 \\
-\frac{4 \mu}{3} \frac{m}{\rho^2} & \frac{4 \mu}{3 \rho} & 0 & 0 & 0 & 0 \\
-\frac{\mu n}{\rho^2} & 0 & \frac{\mu}{\rho} & 0 & 0 & 0 \\
-\frac{4 \mu}{3} \frac{m^2}{\rho^3} - \frac{\mu n^2}{\rho^3} & \frac{4 \mu}{3} \frac{m}{\rho^2} & \frac{\mu n}{\rho^2} & 0 & 0 & 0 \\
0 & 0 & 0 & 0 & 0 & 0 \\
0 & 0 & 0 & 0 & 0 & 0 \\
\end{bmatrix}
\end{equation}
\begin{equation}
\frac{\partial W_2}{\partial U} = 
\begin{bmatrix}
0 & 0 & 0 & 0 & 0 & 0 \\
-\mu \left ( \frac{\partial_y m}{\rho^2} - \frac{2 m \partial_y \rho}{\rho^3} \right ) & - \mu \frac{\partial_y \rho}{\rho^2} & 0 & 0 & 0 & 0 \\
-\frac{4 \mu}{3} \left ( \frac{\partial_y n}{\rho^2} - \frac{2 n \partial_y \rho}{\rho^3} \right ) & 0 & -\frac{4 \mu}{3} \frac{\partial_y \rho}{\rho^2} & 0 & 0 & 0 \\
w_1 & w_2 & w_3 & 0 & 0 & 0 \\
0 & 0 & 0 & 0 & 0 & 0 \\
0 & 0 & 0 & 0 & 0 & 0 \\
\end{bmatrix}
\end{equation}\\
where $w_1 = -\frac{4 \mu}{3} \left (\frac{2 n \partial_y n}{\rho^3} - \frac{3 n^2 \partial_y \rho}{\rho^4}  \right ) - \mu \left (\frac{2 m \partial_y m}{\rho^3} - \frac{3 m^2 \partial_y \rho}{\rho^4}  \right )$, $w_2 = \mu \left ( \frac{\partial_y m}{\rho^2} - \frac{2 m \partial_y \rho}{\rho^3}  \right )$, $w_3 = \frac{4 \mu}{3} \left ( \frac{\partial_y n}{\rho^2} - \frac{2 n \partial_y \rho}{\rho^3}  \right )$.\\
\begin{equation}
\frac{\partial W_2}{\partial U_x} = 
\begin{bmatrix}
0 & 0 & 0 & 0 & 0 & 0 \\
-\mu \frac{m}{\rho^2} & \frac{\mu}{\rho} & 0 & 0 & 0 & 0 \\
-\frac{4 \mu}{3} \frac{n}{\rho^2} & 0 & \frac{4 \mu}{3} \frac{1}{\rho} & 0 & 0 & 0 \\
-\frac{4 \mu}{3} \frac{n^2}{\rho^3} - \frac{\mu m^2}{\rho^3} & \mu \frac{m}{\rho^2} & \frac{4 \mu}{3}\frac{ n}{\rho^2} & 0 & 0 & 0 \\
0 & 0 & 0 & 0 & 0 & 0 \\
0 & 0 & 0 & 0 & 0 & 0 \\
\end{bmatrix}
\end{equation}

\section*{•}{References}

\bibliography{mybibfile}

\end{document}